\documentclass[11pt]{article}
\pdfoutput=1

\usepackage{arxiv}

\usepackage[utf8]{inputenc} 
\usepackage[T1]{fontenc}    
\usepackage{hyperref}       
\usepackage{url}            
\usepackage{booktabs}       
\usepackage{amsfonts}       
\usepackage{nicefrac}       
\usepackage{microtype}      
\usepackage{lipsum}		
\usepackage{graphicx}
\usepackage[numbers]{natbib}
\usepackage{doi}
\usepackage{subfiles}
\usepackage{braket}
\usepackage{amsmath,amsfonts,amssymb}

\title{Trapped atoms in spatially-structured vector light fields}

\author{ \href{https://orcid.org/0000-0002-5363-1194}{\includegraphics[scale=0.06]{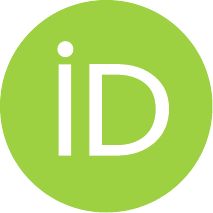}\hspace{1mm}Maurizio Verde} \\
	Institut für Physik\\
	Johannes Gutenberg-Universität\\
	Mainz, 55128, Germany \\
	\texttt{mauverde@uni-mainz.de} \\
	\And
	\href{https://orcid.org/0000-0001-5341-7860}{\includegraphics[scale=0.06]{orcid.pdf}\hspace{1mm}Ulrich Poschinger} \\
	Institut für Physik\\
	Johannes Gutenberg-Universität\\
	Mainz, 55128, Germany \\
 \And
	\href{https://orcid.org/0000-0002-5697-2568}{\includegraphics[scale=0.06]{orcid.pdf}\hspace{1mm}Ferdinand Schmidt-Kaler} \\
	Institut für Physik\\
	Johannes Gutenberg-Universität\\
	Mainz, 55128, Germany \\
 \And
 	\href{https://orcid.org/0000-0002-1153-7124}{\includegraphics[scale=0.06]{orcid.pdf}\hspace{1mm}Christian T. Schmiegelow} \\
	Universidad de Buenos Aires, Facultad de Ciencias Exactas y Naturales, Departamento de Física. Buenos Aires, Argentina,\\
	CONICET - Universidad de Buenos Aires, Instituto de Física de Buenos Aires (IFIBA). Buenos Aires, Argentina\\
}

\begin{document}
\flushbottom
\maketitle
\begin{abstract}
Spatially-structured laser beams, eventually carrying orbital angular momentum, affect electronic transitions of atoms and their motional states in a complex way. We present a general framework, based on the spherical tensor decomposition of the interaction Hamiltonian, for computing atomic transition matrix elements for light fields of arbitrary spatial mode and polarization structures. We study both the bare electronic matrix elements, corresponding to transitions with no coupling to the atomic center-of-mass motion, as well as the matrix elements describing the coupling to the quantized atomic motion in the resolved side-band regime. 
We calculate the spatial dependence of electronic and motional matrix elements for tightly focused Hermite-Gaussian, Laguerre-Gaussian and for radially and azimuthally polarized beams. We show that near the diffraction limit, all these beams exhibit longitudinal fields and field gradients, which strongly affect the selection rules and could be used to tailor the light-matter interaction. 
The presented framework is useful for describing trapped atoms or ions in spatially-structured light fields and therefore for designing new protocols and setups in quantum optics, -sensing and -information processing.
\end{abstract}

%
%
\thispagestyle{empty}

\section*{Introduction}

Most contemporary experiments in the domain of quantum optics and related fields are based on the precise control of matter by light at the single-atom level. The underlying techniques and devices have seen a considerable development throughout the last decades, leading to many emerging applications, such as atomic frequency standards, sensors for magnetic or electric fields, rotations or inertial forces, quantum simulators and quantum computers.  On the other hand, \textit{structured} laser beams, featuring superior focusing properties, have already been successfully applied in many fields of physics and technology, including optical trapping~\cite{kuga1997novel} and manipulation of microscopic particles~\cite{he1995direct,padgett2011tweezers,stilgoe2022controlled} as well as to superresolution microscopy techniques~\cite{hell1994breaking,swartzlander2001peering,furhapter2005spiral}. In the context of quantum technology, other applications, such as atomic clocks~\cite{sanner2019optical,lange2021improved} and quantum computers~\cite{haffner2008quantum,henriet2020quantum}, may also benefit from tailoring light-matter interactions beyond the options offered by commonly used Gaussian beams. A generalized theoretical framework of light-matter interaction for general structured laser beams is needed to proceed beyond specific case studies such as those for Bessel- or Lauguerre-Gaussian \textit{vortex} beams~\cite{schmiegelow2012light,Solyanik2019,PhysRevA.107.023106}. Modern optics instrumentation allows for generating and characterizing a rich variety of different vortex beams with complex polarization structure~\cite{padgett2017orbital}, or tightly focused light well beyond the validity range of the paraxial approximation~\cite{QUABIS20001,PhysRevA.79.033830,alber2017focusing,araneda2020panopticon}. These tailored light fields are generated by spatial light modulators which allow for programmable and switchable beams, or by holographic plates  which allow for very precise control over the spatial structure of laser fields. 
On the other hand, many experimental platforms can provide well localized single atomic absorbers. Examples include single ions or ion crystals in Paul traps~\cite{jefferts1995paul} or Penning traps~\cite{blaum2020perspectives}, or neutral atoms in optical lattices or tweezer arrays~\cite{kaufman2012cooling}, as well as solid state systems like rare-earth ions in a host crystal~\cite{kolesov2012optical}, or nitrogen or silicon-color centers in diamond crystals~\cite{bradac2019quantum}. 

In this work we present a semiclassical treatment of light-matter interaction, with quantized matter and classical electromagnetic fields, beyond the typically made two assumptions, that the atomic absorbers are strongly localized as compared to the spatial variation scales of electromagnetic fields, and that light can be described by plane-waves or by Gaussian beams within the paraxial approximation. Specifically, we consider a single trapped ion as \textit{partially} localized, with its electronic state highly localized near the nucleus within approximately 0.1~nm and its center-of-mass wave function spreading a few tens of nanometers. The latter scale approaches, but it is still smaller than the wavelengths of driving light fields for typical electric dipole transitions. The results naturally extend to trapped neutral atoms, if the confining potential is sufficiently strong. We do not distinguish between atomic ions and neutral atoms, assuming strong confinement in both cases. On the other side, focused beams are described by accounting for their inherent vector character beyond the paraxial approximation. From these requirements, we derive a theoretical framework which allows for predicting, designing and harnessing the properties of structured light fields. Our work is supposed to serve also as a tool to explore assets when applying tailored light fields to localized atomic samples or condensed matter absorbers. We provide the code which was used to generate all results shown in this work. Earlier versions of this code were used by us to predict and simulate the expected response of trapped ions in various experiments using structured beams; such as in the demonstration of the transfer of optical angular momentum to a single ion~\cite{schmiegelow2016transfer}, in the super-resolution microscopy of ion wave packets~\cite{drechsler2021optical} and in the demonstration of the coherent excitation of motional states with vortex beams~\cite{stopp2022coherent}. This manuscript is also designed as a guide for using our code, see the \href{https://community.wolfram.com/groups/-/m/t/3099859}{Mathematica Notebook}, such that readers can adapt it to their specific use-cases.

We start by presenting a closed form to evaluate the focal electric fields of propagating waves beyond the paraxial approximation and taking into account the full vector properties of the field. Then, we discuss the interaction Hamiltonian, taking the electric dipole (E1) and electric quadrupole (E2) transitions into account, and discuss several interesting example scenarios, with special attention on vortex light carrying orbital angular momentum. We describe arbitrary light-matter interaction geometries represented by different angles between the quantizing magnetic field and the beam's propagation direction and we account for multiple polarization distributions by using the spherical tensor decomposition of the interaction Hamiltonian and several mathematical identities. For example, we show that for strongly focused beams, the polarization in the focal plane is in general position-dependent, which leads to complex spatial dependencies of the excitation probabilities. Evidence of this property was experimentally demonstrated in our previous works with ions centered on vortex beams~\cite{schmiegelow2016transfer,Quinteiro2017}. Moreover, we demonstrate how more complex features arise if the spatial position of the atomic absorber is varied transversally across the beam~\cite{Solyanik2019,Afanasev_2018}. This can be either realized by scanning the ion's position across the beam or by moving the beam with respect to a fixed absorber~\cite{drechsler2021optical,PhysRevLett.127.263603}. Finally, we extend the theoretical framework accounting for the atomic center-of-mass motion, which is fundamental to describe tightly bound neutral atoms in optical tweezers, or trapped ions in Paul traps. In these cases, the quantized external degrees of freedom couple to the electric field via light forces~\cite{stopp2022coherent}. In the last section, we sketch  applications for tailored light fields, with emphasis on quantum computing and -simulation. We point out that structured light may provide an addressable way to control and measure motion transverse to the beam's propagation direction, improve cooling of atomic sensors and mitigating frequency shift from parasitic transitions for atomic clocks~\cite{PhysRevLett.129.253901}. Incentivized by these perspectives, we expect further applications or generalizations of the presented framework.

\section*{Spatially structured light fields}

Propagating light is typically described by plane waves or paraxial waves. Plane waves are exact  solutions of Maxwell's equations, which have infinite spatial extent and are therefore unphysical. Paraxial waves are approximate solutions of Maxwell's equations, which more properly describe light fields encountered in realistic settings. The paraxial approximation, as derived e.g. in \cite{lax1975maxwell,siegman1986lasers}, relies on expanding the electromagnetic field in terms of powers of the factor $\frac{k_r}{k_z}$, where $k_r$ and $k_z$ denote, respectively, the transversal and longitudinal wavenumbers. Usually, in describing the light-matter interaction, only the zeroth order is retained, which accounts for transversely polarized propagating fields. Typical paraxial solutions are the ubiquitous Gaussian-type beams, i.e. the Hermite-Gauss (HG) and Laguerre-Gauss (LG) solutions, where the paraxial factor can be expressed as $(k_z w_0)^{-1}$, with $k_z=\frac{2\pi}{\lambda}$ being the wavenumber and $w_0$ the beam waist at the focus. These solutions accurately account for light beams where the transverse mode profile slowly changes along the propagation direction. In particular, LG solutions adequately describe light fields carrying orbital angular momentum, associated to helical wavefronts, when the beams are softly focused. In this regime, the orbital angular momentum is independent from the polarization degree of freedom, i.e. the spin angular momentum. However, even if these two quantities have been demonstrated to be separately gauge invariant~\cite{barnett2001optical}, they cannot be associated with angular momentum operators obeying the standard commutation relations~\cite{van1994commutation,van1994spin} and in general they cannot be properly decoupled. As discussed e.g. in  \cite{bliokh2015spin,barnett1994orbital,takenaka1985propagation}, the distinction between orbital and spin angular momentum breaks down for strongly confined fields close to the diffraction limit \cite{novotny2012principles,dorn2003focus,zhan2013vectorial}, for beams featuring a spatially modulated polarization, and for light fields near the singularity of optical vortices \cite{quinteiro2015formulation,quinteiro2017formulation}. In these cases, the non-separability between spin and orbital angular momentum must be taken into account. We refer to these fields as spatially structured vector light fields and provide a rigorous framework for computing transition matrix elements for such fields, valid beyond the paraxial approximation.

Propagation of vector light and its properties in the focus have been extensively studied and predict two main features. First, spatial variations which are strong on the wavelength scale lead to the appearance of longitudinal electric field components. This fact can be seen as a consequence of the integral representation of focal fields introduced in~\cite{richards1959electromagnetic}. Indeed, when beams are described as a superposition of plane waves propagating along different directions, a unique k-vector is intrinsically ill-defined and longitudinal electric fields accordingly emerge. We stress here, that such fields are completely different from the pure longitudinal electromagnetic fields. Second, the separation between polarization and spatial degrees of freedom is no longer valid and vector solutions involving non-separable combinations of spatial and polarization modes become necessary \cite{chen2002analyses,wang2020vectorial}. Here, we discuss this aspect, with special attention to the case of LG beams, explicitly showing the non-separability between spin and orbital angular momentum, whose interplay can generate strong longitudinal fields along the propagation ($z$) axis \cite{huang2011vector,forbes2021relevance,monteiro2009angular}. 

\begin{figure}[ht!]
\centering
\includegraphics[scale = 0.25]{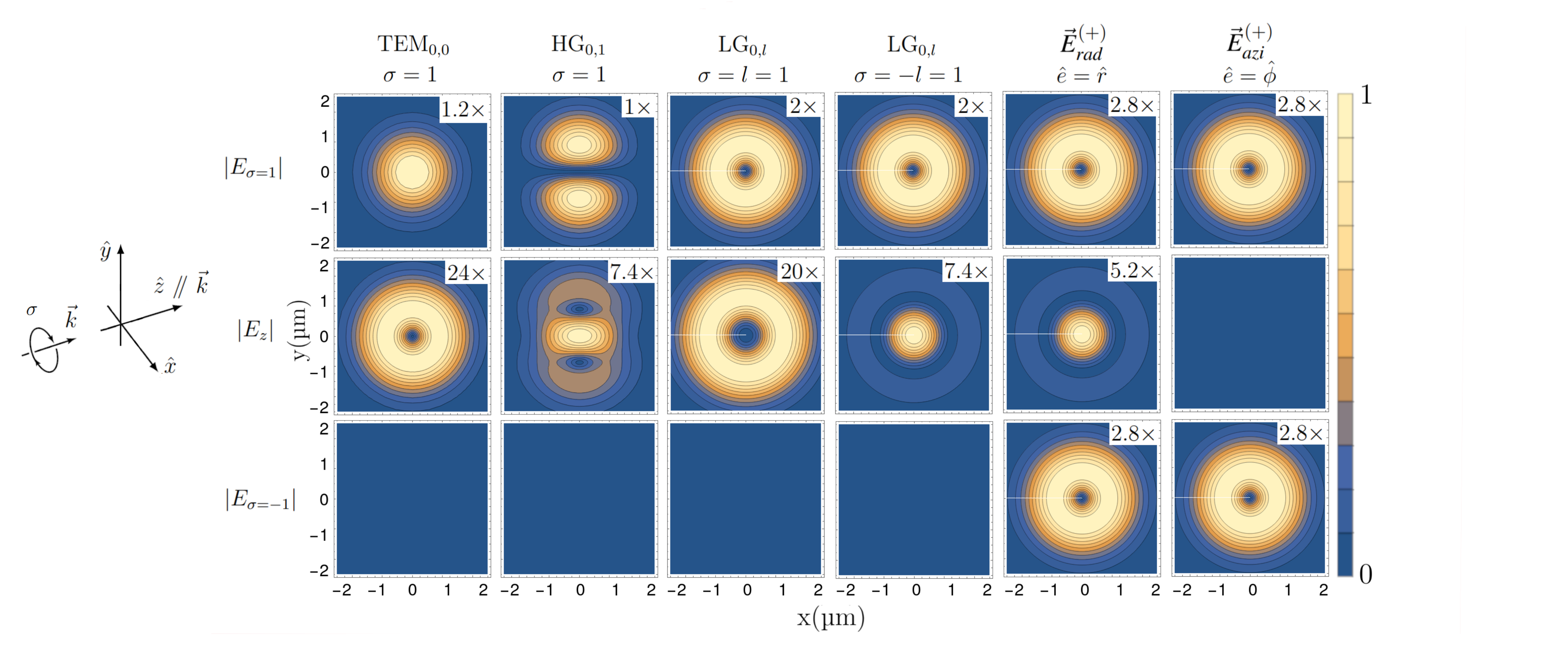}
\caption{Normalized transverse profiles of the electric field modulus for longitudinal and circular polarization components, calculated within the \href{https://community.wolfram.com/groups/-/m/t/3099859}{Mathematica Notebook}. Here and in all following plots the scale factor in panel indicates by how much the shown values are scaled up to cover the full color scale. Large factors correspond to weak polarization components. Different vector beam types are shown, e.g. HG and LG beams as described by equations~(\ref{eq:Hermite-Gaussian-expression}) and (\ref{eq:Laguerre-Gaussian-expression}) and radially and azimuthally polarized beams described by equations~(\ref{eq:radial-azimuthal-vector-beams}). All the vortex beams pertain to radial index $p=0$ and orbital angular momentum $l=\pm1$. All the beams are close to the diffraction limit with a chosen beam waist of $w_0=1\; \mu m$ for the optical wavelength is $\lambda=0.729\; \mu m$. }.
\label{fig:1}
\end{figure}

Analytical expressions of electric fields have been carried out in complete compliance with Maxwell’s equations as non-paraxial corrections of the paraxial solution~\cite{davis1979theory}. Focal fields can therefore be explicitly calculated to the desired degree of accuracy. We start from the plane polarized Lorentz-gauge vector potential $\vec{A}(\vec{r},t)$ describing an electromagnetic beam in the free space, with no charges or currents, propagating along the $z$ axis at wavenumber $k$ and optical frequency $\omega$:
\begin{equation}\label{eq:vector-potential}
    \vec{A}(\vec{r},t) = f(\vec{r})(\vec{e}_x+i\sigma\vec{e}_y)e^{i (k z-\omega t)} + c.c.,
\end{equation}
where $f(\vec{r})$ is a slowly varying function obeying the Helmholtz equation. We use the Lorentz condition $\phi=(\frac{i}{k})\nabla\cdot\Vec{A}$ to calculate the electric field from the vector potential alone $\vec{E}(\vec{r},t) = -(\frac{i}{k})\nabla(\nabla\cdot\Vec{A}(\vec{r},t))-i k \Vec{A}(\vec{r},t)$. As shown in~\cite{quinteiro2017formulation}, being $\vec{e}_i$ the unit vectors along the directions $i=x,y,z$, we reach the general expression:
\begin{equation} \label{eq:vector-field}
\vec{E}^{(+)}(\vec{r},t) = \left\{(\vec{e}_x+i\sigma\vec{e}_y)f(\vec{r})+\frac{i}{k}[\partial_x f(\vec{r})+i\sigma\partial_y f(\vec{r})]\vec{e}_z\right\}e^{i (k z-\omega t)},
\end{equation}
where we neglected the higher order corrections proportional to the second order spatial derivatives $\partial_i\partial_jf(\vec{r})$ and we considered the positive frequency part only. The ellipticity parameter $\sigma$ describes the optical polarization. In the soft-focusing limit of vanishing  gradients of the mode function, $\sigma=0$ describes purely linear polarization along $x$, while $\sigma=\pm 1$ describes purely circular polarizations. The two peculiar properties of vector beams explicitly appear in equation~(\ref{eq:vector-field}). Longitudinal field components $\propto \vec{e}_z$ are proportional to transverse gradients of the mode function $\partial_{x,y}f(\vec{r})$. Moreover, these components and therefore the overall polarization depend on the position within the light field. To account for tightly focused HG beams, we use equation~(\ref{eq:vector-field}) by means of the mode function $f^{HG}_{n,m}(\vec{r})$:
\begin{equation} \label{eq:Hermite-Gaussian-expression}
f^{HG}_{n,m}(x,y,z) = \frac{E_0}{\sqrt{\pi}}\frac{w_0}{w(z)}H_n\left(\frac{\sqrt{2}x}{w(z)}\right)H_m\left(\frac{\sqrt{2}y}{w(z)}\right) \,\exp\left(-\frac{x^2+y^2}{w(z)^2}+i k z\frac{x^2+y^2}{2 R(z)}+i \psi(z)\right),
\end{equation}
where $w_0$ is the beam waist, $H_n$ and $H_m$ are the Hermite polynomials and $\psi(z)=\arctan(z/z_R)$ is the usual Gouy phase for the Rayleigh length $z_R=kw_0^2/2$. In particular, the mode function $f^{HG}_{0,0}(\vec{r})$ within equation~(\ref{eq:vector-field}) allows to deal with highly focused TEM$_{00}$ Gaussian beams. Similarly, focused LG beams are described by the mode function $f^{LG}_{l,p}(\vec{r})$, which depends on the integer radial index $p\geq0$ and the integer azimuthal index $l$ associated to orbital angular momentum.
In cylindrical coordinates~$\{\rho,\phi,z\}$, the mode function $f^{LG}_{l,p}(\vec{r})$ reads: 
\begin{equation} \label{eq:Laguerre-Gaussian-expression}
f^{LG}_{l,p}(\rho,\phi,z) = \frac{E_0}{\sqrt{2}} C_p^l\frac{w_0}{w(z)}\left(\frac{\rho\sqrt{2}}{w(z)}\right)^{|l|}L_p^{|l|}\left(\frac{2\rho^2}{w(z)^2}\right) \,\exp\left(i l \phi-\frac{\rho^2}{w(z)^2}+i k z\frac{\rho^2}{2 R(z)}+i \psi_{LG}(z)\right),
\end{equation}
where $C_p^l=\sqrt{2 p!/\pi (p + |l|)!}$ is a normalization constant, $L_p^{|l|}(x)$ are the generalized Laguerre polynomials and $\psi_{LG}(z)=-(2p+|l|+1)\arctan(z/z_R)$ is the generalized Gouy phase. 
Other vortex fields, sharing the same transverse intensity distribution of LG beams, but a different polarization distribution, are the radially and azimuthally polarized vector beams. They are superpositions of LG beams:
\begin{equation} \label{eq:radial-azimuthal-vector-beams}
\begin{split}
\vec{E}^{(+)}_{rad} & =\frac{1}{\sqrt{2}}\left[ \vec{E}^{(+)}_{LG}(1,0,1)+\vec{E}^{(+)}_{LG}(-1,0,-1) \right]\\
\vec{E}^{(+)}_{azi} & =\frac{1}{\sqrt{2}}\left[\vec{E}^{(+)}_{LG}(1,0,1)-\vec{E}^{(+)}_{LG}(-1,0,-1)\right].
\end{split}
\end{equation}
These two beams display different properties in the focal plane \cite{maurer2007tailoring}. In Fig.~\ref{fig:1} we plot the transverse profiles in the focal plane for the longitudinal and circular field components 
\begin{equation}\label{eq:field-components}
E_z(x,y)=\vec{E}(x,y) \cdot \vec{e}_z \qquad E_{\sigma=\pm 1}(x,y)=\vec{E}(x,y) \cdot (\vec{e}_x+i\sigma \vec{e}_y)
\end{equation}
for several tightly focused beams. One can recognize some of the most prominent features: Spatially modulated longitudinal electric field components appear for all beams except for the azimuthally polarized vortex light, which displays a perfectly vanishing longitudinal electric field component. The radially polarized one shows a distinctive longitudinal field concentrating at its center on the focus. Moreover, the LG beam with orbital angular momentum $l=1$ and aligned polarization $l=\sigma=1$ displays a longitudinal field with a doughnut-shaped profile, while the anti-aligned LG beam, $l=-\sigma=1$ , displays a Gaussian-shaped longitudinal field which "fills" the vortex center. 

\section*{Spherical tensor decomposition for light-matter interaction}

Here, we present a framework which can be used to compute atomic transition matrix elements for general vector light fields of arbitrary mode structure. Specifically, it can be applied to electric dipole and electric quadrupole transitions for atomic species with or without nuclear spin.\\
We consider an atom, with a single valence electron at position $\vec{r}=(x,y,z)$ with respect to the nucleus located at $\vec{R}=(X,Y,Z)$. The atom is interacting with a monochromatic laser beam, with an optical frequency $\omega$ and an arbitrary spatial mode structure $\vec{E}(\vec{R}+\vec{r},t) = \vec{E}^{(+)}(\vec{R}+\vec{r}) + c.c. $ characterized by the appearance of both transversal and longitudinal components and a general polarization distribution. By considering the first two terms in the multipole expansion of the electric field $\vec{E}^{(+)}(\vec{R}+\vec{r})$ around the nuclear position $\vec{R}$, we can write the interaction Hamiltonian that describes optically driven electric dipole and quadrupole transitions, as follows: 
\begin{equation} \label{interaction hamiltonian}
\hat{H}_{I}^{(+)}(\vec{r};\vec{R}) =\left.\sum_{i}\hat{r}_iE_i^{(+)}\right|_{\vec{\hat{R}}}+\left.\sum_{i,j}\hat{r}_i\hat{r}_j\frac{\partial E_j^{(+)}}{\partial r_i}\right|_{\vec{\hat{R}}}+\mathcal{O}(r^3)
\end{equation}
the electric field and its gradients are evaluated at the atom's center-of-mass position $\vec{R}$. The first and second terms describe the dipole and the quadrupole interaction, respectively. This approximation is accurate as long as the valence electron's spatial wavefunction is strongly localized on the scale of the variations of the electric field. As shown in \cite{van1994selection,sonnleitner2017rontgen}, two further higher order terms, one accounting for interactions with two electrons and the nucleus and the so-called Röntgen term, are by several orders of magnitude smaller. The generalization of this treatment to magnetic transitions or to higher-order terms in the electric field multipole expansion, which play a role in optical frequency standards~\cite{PhysRevLett.129.253901,peshkov2023excitation}, is straightforward. If the optical frequency lies near one specific electronic resonance, we use the rotating wave approximation, i.e. we neglect the oscillatory time-dependence and only consider the positive-frequency part $\hat{H}^{(+)}_{I}$. Each term in $\hat{H}^{(+)}_{I}$ involves a combination of the electronic position operator components, multiplying the electric field components or their gradients. We can then rewrite all these terms, using a more suitable set of electronic operators, to take advantage of the symmetry properties of the electronic eigenstates.

For atoms, the spherical symmetry of the nucleus' Coulomb potential leads to electronic eigenstates factorizing in an angular part, described by universal spherical harmonics $Y_{J, m}(\theta,\phi)$, and a radial part $\Phi_{n,J}(r)$, where $n$ is the principal quantum number, $J$ is the total angular momentum eigenvalue and $m$ its projection on the magnetic quantization axis. To separate the radial and angular part of $\hat{H}^{(+)}_{I}$ in the calculation of transition matrix elements between different electronic eigenstates, a natural choice is the use of irreducible spherical tensor operators. These have well-defined transformation properties under rotations. Their matrix elements, in the basis of angular momentum eigenstates, are written in terms of Clebsch-Gordan coefficients by means of the Wigner-Eckart theorem~\cite{sakurai1994modern}. Here, as we consider electronic dipole and quadrupole transitions, we rewrite $\hat{H}^{(+)}_{I}$ by using rank-1 and rank-2 irreducible spherical tensors $\hat{T}_{\varDelta m}^{\varDelta J}$, where $\varDelta J=0,1,2$ is the change of total angular momentum and $\varDelta m=+\varDelta J,\cdots,-\varDelta J$ is the change of its projection on the quantization axis. Without loss of generality, we assume that the quantization axis is induced by an external magnetic field $\Vec{B}=B_0\hat{z}$ aligned along the $z$-direction. Here, it is worth noticing that even if the direction of the quantization axis is completely arbitrary, the introduction of a magnetic-induced quantization axis is relevant for two reasons. First, an external magnetic field allows to define a common quantization axis, independently on the polarization of the driving laser beams. Secondly, thanks to the Zeeman splitting the different magnetic transitions can be separately laser driven. Given this choice, the position operators, in terms of spherical tensors, are: 
\begin{equation}
\hat{x} = \hat{r}\sqrt{\frac{2\pi}{3}}(\hat{T}^{1}_{1} + \hat{T}^{1}_{-1}) \qquad \hat{y} = -i\hat{r}\sqrt{\frac{2\pi}{3}}(\hat{T}^{1}_{1} - \hat{T}^{1}_{-1}) \qquad \hat{z} =  \hat{r}\sqrt{\frac{4\pi}{3}}\hat{T}^{1}_{0},
\end{equation}
and the dipole interaction Hamiltonian reads: 
\begin{equation} \label{eq:dipole-Hamiltonian}
\begin{split}
\hat{H}^{(+)}_{I,E1}  = \sqrt{\frac{2\pi}{3}}\hat{r}&[\hat{T}^{1}_{1}\big( E_x - i E_y \big) +  \sqrt{2}\hat{T}^{1}_{0}E_z + \hat{T}^{1}_{-1}\big( E_x + i E_y \big)], 
\end{split}
\end{equation}
where the electric field components $E_i$ are evaluated in the center-of-mass position and do not rely on electronic coordinate operators. From equation~(\ref{eq:dipole-Hamiltonian}), the relative strengths for electric dipole transitions characterized by $\Delta m=0,\pm1$ can be evaluated. In order of doing that, the propagation direction $\vec{k}$ of the light field has to be chosen with respect to magnetic quantization axis. For instance, if $\vec{k}$ is parallel to $\vec{B}$, then $E_z(\vec{r})$ describes the longitudinal field component, while $E_x(\vec{r})$ and $E_y(\vec{r})$ are the transverse components. In that case, $\Delta m=0$ are driven by the longitudinal component field $E_z(\vec{r})$, and $\Delta m=\pm1$ transitions are driven by the transverse components, namely by the circular polarization components $E_x(\vec{r})\mp i E_y(\vec{r})$. The relative transition strengths, are proportional to the electric field components described by spatial mode profiles shown in Fig.~\ref{fig:1}.\\
To obtain the quadrupole interaction part in terms of the spherical tensor operators, we express the terms quadratic in the position operators by exploiting the algebra of irreducible spherical tensors
\begin{equation}
\hat{T}^{J_1}_{m_1}\hat{T}^{J_2}_{m_2} = \sum_{J,m}\hat{T}^{J}_{m}\braket{J,m|J_1,m_1;J_2,m_2}
\end{equation}
with the Clebsch-Gordan coefficients $\braket{J,m|J_1,m_1;J_2,m_2}$, as follows:
\begin{alignat*}{3} \label{eq1}
\hat{x}^2 & = \hat{r}^2\frac{2\pi}{3}\left(\hat{T}^{2}_{2} + \frac{2}{\sqrt{3}}\hat{T}^{0}_{0} + \frac{2}{\sqrt{6}}\hat{T}^{2}_{0} + \hat{T}^{2}_{-2}\right) \qquad
&& \hat{x}\hat{z} && = \hat{r}^2\sqrt{2}\frac{2\pi}{3}(\frac{1}{\sqrt{2}}\hat{T}^{1}_{1} + \frac{1}{\sqrt{2}}\hat{T}^{2}_{1} - \frac{1}{\sqrt{2}}\hat{T}^{1}_{-1} + \frac{1}{\sqrt{2}}\hat{T}^{2}_{-1}) \\
\hat{y}^2 & = -\hat{r}^2\frac{2\pi}{3}\left(\hat{T}^{2}_{2} - \frac{2}{\sqrt{3}}\hat{T}^{0}_{0} - \frac{2}{\sqrt{6}}\hat{T}^{2}_{0} + \hat{T}^{2}_{-2}\right) \qquad
&&\hat{z}\hat{x} && = \hat{r}^2\sqrt{2}\frac{2\pi}{3}(-\frac{1}{\sqrt{2}}\hat{T}^{1}_{1} + \frac{1}{\sqrt{2}}\hat{T}^{2}_{1} + \frac{1}{\sqrt{2}}\hat{T}^{1}_{-1} + \frac{1}{\sqrt{2}}\hat{T}^{2}_{-1})\\
\hat{z}^2 & =  \hat{r}^2\frac{4\pi}{3}\left(-\frac{1}{\sqrt{3}}\hat{T}^{0}_{0} + \sqrt{\frac{2}{3}}\hat{T}^{2}_{0}\right) \qquad
&&\hat{y}\hat{z} && = -i\hat{r}^2\sqrt{2}\frac{2\pi}{3}\left(\frac{1}{\sqrt{2}}\hat{T}^{1}_{1} + \frac{1}{\sqrt{2}}\hat{T}^{2}_{1} + \frac{1}{\sqrt{2}}\hat{T}^{1}_{-1} - \frac{1}{\sqrt{2}}\hat{T}^{2}_{-1}\right)\\
\hat{x}\hat{y} & =  -i\hat{r}^2\frac{2\pi}{3}(\hat{T}^{2}_{2} - \frac{2}{\sqrt{2}}\hat{T}^{1}_{0} - \hat{T}^{2}_{-2}) \qquad
&&\hat{z}\hat{y} && = -i\hat{r}^2\sqrt{2}\frac{2\pi}{3}(- \frac{1}{\sqrt{2}}\hat{T}^{1}_{1} + \frac{1}{\sqrt{2}}\hat{T}^{2}_{1} - \frac{1}{\sqrt{2}}\hat{T}^{1}_{-1} - \frac{1}{\sqrt{2}}\hat{T}^{2}_{-1}).\\
\hat{y}\hat{x} & =  -i\hat{r}^2\frac{2\pi}{3}(\hat{T}^{2}_{2} + \frac{2}{\sqrt{2}}\hat{T}^{1}_{0} - \hat{T}^{2}_{-2})
\end{alignat*}
We neglect the terms containing the spherical tensor $\hat{T}^{0}_{0}$ as it does not mediate neither electric dipole nor electric quadrupole transitions and it is therefore irrelevant for the following analysis. The terms depending on $\hat{T}^{1}_{\varDelta m}$ with $\varDelta m_1=0,\pm1$ describe electric quadrupole transitions with $\varDelta J=1$. The corresponding Hamiltonian can be written as:

\begin{equation}
\begin{aligned} \label{eq:quadrupole-Hamiltonian-j1}
\hat{H}^{(+)}_{I,E2,\varDelta J=1}  = 
\frac{2\pi}{3}\hat{r}^2&\bigg[\hat{T}^{1}_{1}\big( \partial_x E_z - \partial_z E_x - i \partial_y E_z + i \partial_z E_y \big) +\hat{T}^{1}_{0}\frac{2i}{\sqrt{2}}\big( \partial_x E_y - \partial_y E_x\big)  + \\
&+\hat{T}^{1}_{-1}\big( \partial_z E_x - \partial_x E_z  - i \partial_y E_z + i \partial_z E_y\big)\bigg].
\end{aligned}
\end{equation}

The remaining terms, containing $\hat{T}^{2}_{\varDelta m}$, with $\varDelta m=0,\pm1,\pm2$, describe quadrupole transitions with $\varDelta J=2$ through the interaction Hamiltonian:

\begin{equation}
\begin{aligned} \label{eq:quadrupole-Hamiltonian}
\hat{H}^{(+)}_{I,E2,\varDelta J=2}  = 
\frac{2\pi}{3}\hat{r}^2\bigg[&\hat{T}^{2}_{2}\big( \partial_x E_x - \partial_y E_y - i \partial_x E_y - i \partial_y E_x \big) + \hat{T}^{2}_{1}\big( \partial_x E_z + \partial_z E_x - i \partial_y E_z - i \partial_z E_y \big)  + \\
 +&\sqrt{\frac{2}{3}}\hat{T}^{2}_{0}\partial_z E_z   + \hat{T}^{2}_{-1}\big( \partial_x E_z + \partial_z E_x + i \partial_y E_z + i \partial_z E_y \big) \\
 + &\hat{T}^{2}_{-2}\big( \partial_x E_x - \partial_y E_y + i \partial_x E_y + i \partial_y E_x \big) \bigg],
\end{aligned}
\end{equation}
where we used $\nabla\cdot\vec{E}=0$ to simplify the factor multiplying the tensor component $\hat{T}^{2}_{0}$. All the electric field gradients terms $\partial_i E_j$ are scalar quantities evaluated at the atomic center-of-mass position. Equations~(\ref{eq:dipole-Hamiltonian}),~(\ref{eq:quadrupole-Hamiltonian-j1}) and (\ref{eq:quadrupole-Hamiltonian}), rely on all the electric field components and their gradients. The description of light-matter interaction beyond the paraxial approximation becomes then natural and longitudinal fields or spatially modulated polarization can be easily taken into account. Therefore, writing the interaction Hamiltonians in terms of tensor operators offers a clear advantage when evaluating transition matrix elements for arbitrary vector light fields. \\
The usage of tensor operators allows for invoking the Wigner-Eckart theorem, stating that the matrix element of a tensor operator between two atomic eigenstates with principal quantum numbers $n_{1,2}$, angular momentum quantum numbers $J_{1,2}$ and magnetic quantum numbers $m_{1,2}$ factorizes as
\begin{equation}
    \braket{n_2,J_2,m_2|\hat{T}^{\Delta J}_{\Delta m}|n_1,J_1,m_1}=\frac{\braket{n_2,J_2||\hat{T}^{\Delta J}||n_1,J_1}}{\sqrt{2J_1+1}}\braket{J_1 m_1 \Delta J \Delta m|J_2 m_2}
\end{equation}
where the reduced matrix element $\braket{n_2,J_2||\hat{T}^{\Delta J}||n_1,J_1}$ depends only on the principal and angular momentum quantum numbers, and the Clebsch-Gordan coefficients  $\braket{J_1 m_1 \Delta J \Delta m|J_2 m_2}$ depend only on the angular momentum and magnetic quantum numbers. 
For a given atomic position $\vec{R}$ in a vector light field and given transition of the atomic species of interest, the transition matrix elements factorize as follows:
\begin{equation} \label{eq:matrix-element}
\braket{n_2,J_2,m_2|\hat{H}^{(+)}_{I}(\vec{R})|n_1,J_1,m_1}=f(n_1,n_2,J_1,J_2)\;\mu(\vec{R};J_2,J_1,m_2,m_2),
\end{equation}
Crucially, $f(n_1,n_2,J_1,J_2)$ depends only on the atomic species and the electronic transition, while the \textit{relative transition strength}
\begin{equation}
    \mu(\vec{R};J_2,J_1,m_2,m_2) = \sum_{\substack{i,j \\ =x,y,z}} \sum_{\Delta m=-\Delta J}^{+\Delta J} \braket{J_1 m_1 \Delta J \Delta m|J_2 m_2} c_{ij}^{(\Delta m)} \partial_i E_j\vert_{\vec{R}}
\label{eq:relTransitionStrength}
\end{equation}
contains the properties of the light field and depends on the atomic position $\vec{R}$ as well as on the sub-transition $m_1 \leftrightarrow m_2$ of interest. The coefficients $c_{ij}^{(\Delta m)}$ can be read off from equations~(\ref{eq:quadrupole-Hamiltonian-j1}) and (\ref{eq:quadrupole-Hamiltonian}). For studying the effects arising from using structured light fields, we are only interested in the relative transition strengths $\mu(\vec{R};J_2,J_1,m_2,m_2)$, which can now readily be evaluated for different field types and electronic transitions, now characterized by $\Delta m=0,\pm1,\pm2$. Moreover, we can describe any interaction geometry as parametrized by the angle between the propagation direction and the magnetic quantization axis. In that case, we keep the axis system for the beam's description, we rotate the atomic quantization axis relative to it, and we finally replace the tensor operators in equations (\ref{eq:quadrupole-Hamiltonian-j1}) and (\ref{eq:quadrupole-Hamiltonian}) by the transformed ones. To account for electronic dipole transitions, the same rules will apply to equation~(\ref{eq:dipole-Hamiltonian}). The transformation is achieved by rotating the spherical tensors around a given axis $\vec{a}$ by the angle $\theta$ using the reduced Wigner (small) $d$-Matrices: 
\begin{equation}\label{eq:tensors-rotation}
\hat{T}^{\Delta J}_{\Delta m}(\vec{r}) \rightarrow \tilde{\hat{T}}^{J}_{\Delta m}(\vec{r}) = \sum_{m'}d_{\Delta m,m'}^{\Delta  J}(\theta,\vec{a})\hat{T}^{\Delta  J}_{m'}(\vec{r})
\end{equation}
for both $\Delta J=1,2$. The explicit form of the Wigner (small) $d$-Matrices is well known~\cite{sakurai1994modern}. \\
Finally, the spatial extent of the center-of-mass wavefunction $\Psi(\vec{R})$ (defined with respect to the trap center $\vec{R}_0$) can lead to non-negligible corrections when the matter wavepacket is too large to be considered as a point-like sensor of the electric field and its gradient, but small enough as compared to the  typical beam's spatial lenghts like the optical wavelenght $\lambda$ or the beam waist w$_0$. It is worth underlying here, that this conditions are fulfilled for a usual trapped ion interacting with a structured laser beam, where the wavefunction  $\Psi(\vec{R})$ spreads over tens of nanometers, while the beam spatial variations are in the hundred nanometers range. Under these assumptions, the beam's optical phase is considered constant over $\Psi(\vec{R})$ and the effective relative transition amplitudes are obtained by averaging the matrix element from equation~(\ref{eq:relTransitionStrength}) over the wavepacket's probability distribution:
\begin{equation}
\bar{\mu}(\vec{R}_0;J_2,J_1,m_2,m_2) = \int \mu(\vec{R}_0+\vec{R};J_2,J_1,m_2,m_2)\vert\Psi(\vec{R})\vert^2 d^3\vec{R},
\end{equation}
where $\vec{R}_0$ is the center of the trapping potential. The averaged $\bar{\mu}(\vec{R}_0)$ can predict the residual excitation appearing where the bare transition matrix element $\mu(\vec{R}_0)$ is vanishing. It happens, for instance, when a single ion is placed in the center of the vortex field. In that case, as shown in \cite{stopp2022coherent}, a residual excitation can be measured and characterized as a function of the wavefunction's extent, which probes the non-vanishing electric fields near the vortex phase-singularity.\\
Note that the above formalism can be equivalently applied to atomic species with nuclear spin. In this case the total angular momentum is $F=J+I$, being $I$ the nuclear spin, and equations~(\ref{eq:dipole-Hamiltonian}) and (\ref{eq:relTransitionStrength}) have to be correspondingly modified by taking $\Delta J\rightarrow \Delta F$ and $\Delta m\rightarrow \Delta m_F$. Moreover, it is worth underlying some key aspects of the introduced framework as compared to previous theoretical works~\cite{afanasev2013off,afanasev2016high,Solyanik2019,afanasev2020polarization}. First, it shows a wide applicability spectrum, which goes beyond the treatment of atomic excitation by light carrying orbital angular momentum, and encompasses a broader range of spatially structured vector beams. Second, even when dealing with vortex light, the particular choice of the beam, commonly in the form of Bessel or LG beams, is not the starting point for the theoretical analysis, but only appears for concrete calculations. Finally, in the context of perturbation theory, by following the ideas presented in~\cite{al2022two}, it could be further generalized to multi-photons transitions.

\section*{Spatial dependence of electric quadrupole transitions in vector light beams}

\begin{figure}[ht!]
\centering
\includegraphics[scale = 0.19]{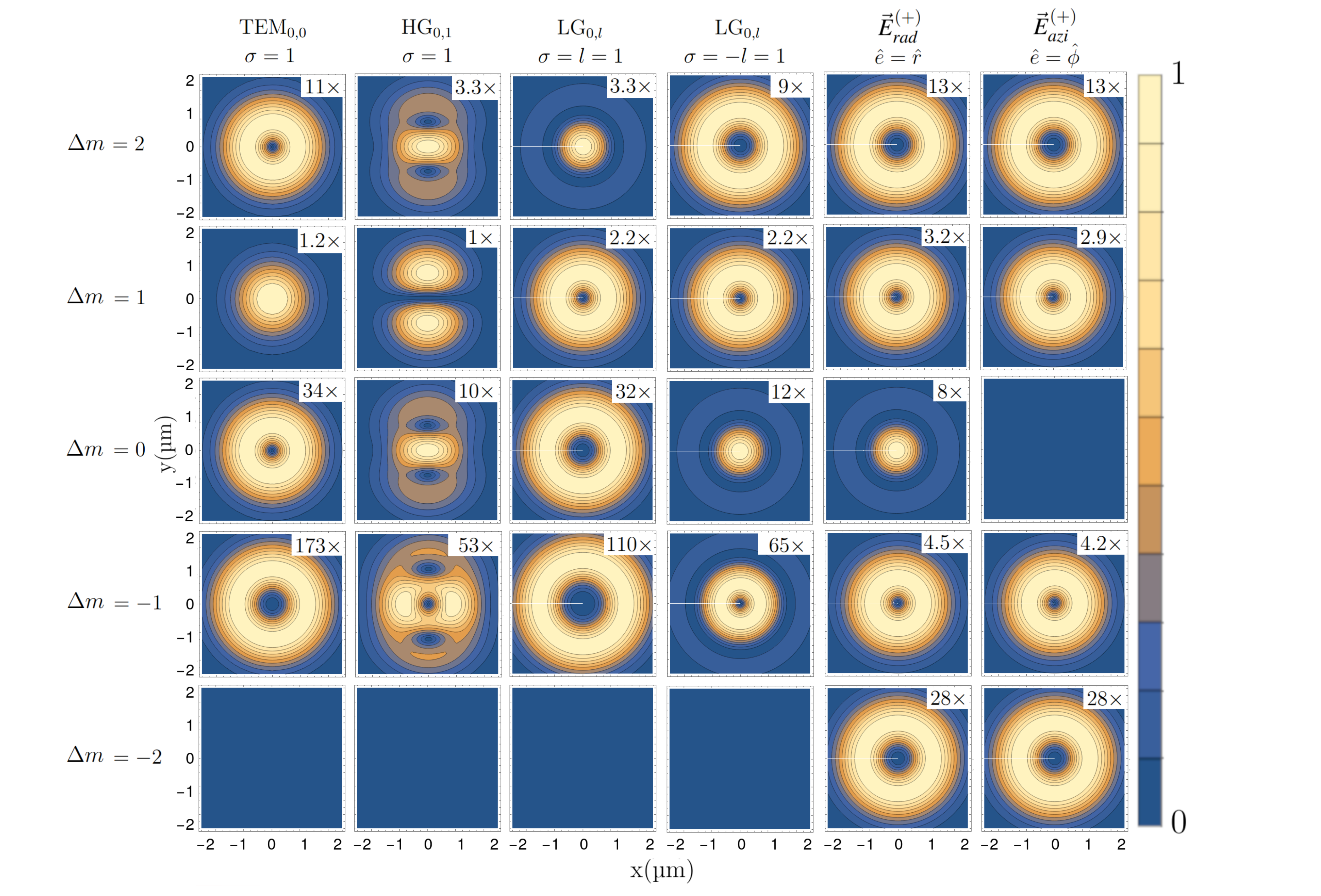}
\caption{Relative transitions strengths computed from equation~(\ref{eq:relTransitionStrength}) for a $\Delta J=2$ electric quadrupole transition. Shown are excitation strength profiles in the focal plane of a drive field propagating along the quantization ($z$) axis, for beam types and parameters as in Fig.~\ref{fig:1}. Profiles are shown for different sub-transitions characterized by the change in the magnetic quantum number $\Delta m$, with $m_1=+1/2$ as the initial state. All the quantities have been calculated within our \href{https://community.wolfram.com/groups/-/m/t/3099859}{Mathematica Notebook}.}.
\label{fig:2}
\end{figure}

In this section, we analyze relative transition strengths for different beam types, sub-transitions and interaction geometries for the particular example of the $\Delta J=2$ electric quadrupole transition $4^2S_{1/2}\leftrightarrow 3^2D_{5/2}$ near 729~nm in a $^{40}$Ca$^+$ ion. The findings presented here directly carry over to other atomic species for identical ratios of beams waist to driving wavelength.\\
In Fig.~\ref{fig:2}, we show transition strengths for sub-transitions between different Zeeman levels, in the focal plane of the various types of structured beams shown in Fig.~\ref{fig:1}, for the case that the drive field propagates along the quantization axis. Here, rotational symmetry around the propagation axis imposes conservation of the $z$-component of total angular momentum. 
Under typical conditions, where the driving wavelength is by far the smallest spatial scale of the light field, mainly the \textit{longitudinal} gradients of transverse components $\partial_z E_x$ and $\partial_z E_y$ drive electric quadrupole transitions. The $\Delta m=\pm 1$ transitions exhibit the strongest coupling, but vanish where the transverse electric field amplitude is zero, as for example on the optical axis of LG and HG beams. However, in the proximity of such field nulls and for tightly focused beams, the strong transversal field gradients drive $\Delta m=\pm 2$ transitions. Furthermore, combinations of transverse and longitudinal gradients $\partial_x E_x$, $\partial_y E_y$ and $\partial_z E_z$ enable driving 
of $\Delta m=0$ transitions. These terms  become dominant in the limit of extremely tight focusing, characterized by strong longitudinal fields.

We recover the experimental findings and theoretical predictions of Refs.~ \cite{schmiegelow2016transfer,Quinteiro2017,Afanasev_2018}. In the particular case of LG beams with a single unit of orbital angular momentum $l=\pm1$, we find the non-vanishing transition strength for $\Delta m=0$ transitions, as occurring in the vortex beam singularity, when the optical spin and orbital angular momenta $\sigma$ and $l$ are anti-parallel, $\sigma=-l$. This behaviour is moreover quantified across the transverse cross-section of the laser beam's focal plane. Transitions with $\Delta m=+1$ are mostly driven by the longitudinal gradient of the electric field and lead to transition strengths proportional to the electric field modulus, which is vanishing on the optical axis of HG and LG beams. Conversely, the transitions strengths for $\Delta m=+2,0$ depend on the transverse electric field gradients. On the optical axis, these vanish for a the Gaussian mode, while assuming maximum values for HG beams. LG beams with topological charge $l=1$, feature strong field gradients on the optical axis, which lead to peak transition strengths in the center for $\Delta m=0$ and vanishing strengths for $\Delta m=+2$. Very small gradients lead to rather weak $\Delta m=-1$ transitions, except for radial and azimuthal light fields, which contain circular polarization $\sigma=-1$ and can also drive strong $\Delta m=-1$ and weak $\Delta m=-2$ transitions. Our formalism, when applied to LG beams, is in full agreement with previous predictions based on different theoretical approaches~\cite{afanasev2013off,afanasev2016high,Solyanik2019,afanasev2020polarization}. 

To explore the effect of different angles $\theta$ between the quantization axis and the propagation direction of the light field, we exploit the rotation properties of spherical tensors. The role of the angle $\theta$ can be intuitively understood for dipole transitions, where the electric field components $E_z$ and $E_{\sigma=\pm1}$, defined via equation~(\ref{eq:field-components}), drive respectively $\Delta m=0$ and $\Delta m=\pm1$ transitions. In that case, rotating the quantization axis effectively means to redefine these components with the consequent modification of matrix element cross-sections. However, in the case of electronic quadrupole transitions, see equations~(\ref{eq:quadrupole-Hamiltonian-j1}) and (\ref{eq:quadrupole-Hamiltonian}), this mapping is much more involved and we cannot follow a simple intuition.

\begin{figure}[ht!]
\centering
\includegraphics[scale = 0.22]{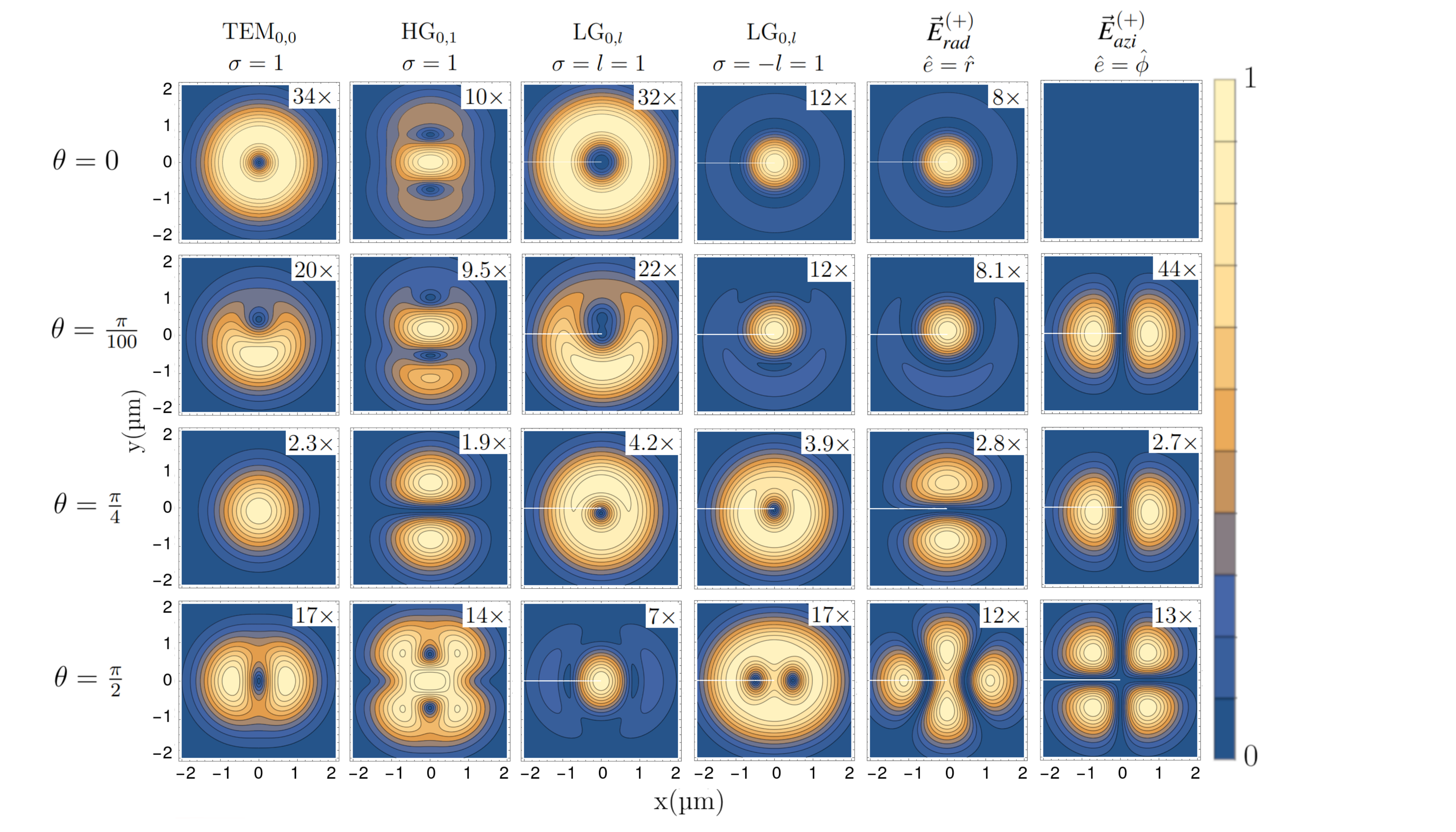}
\caption{Relative transitions strengths computed with our \href{https://community.wolfram.com/groups/-/m/t/3099859}{Mathematica Notebook} from equation~(\ref{eq:relTransitionStrength}) for a $\Delta J=2$ electric quadrupole transition. The beam types represented in Fig.~\ref{fig:1} are used to calculate the excitation strength profiles for a single atom lying in the beam's focal plane. The sub-transition $\Delta m=0$, with initial state $m_1=+1/2$, has been considered for different light-matter interaction geometries which are parametrized by different angles $\theta$, in the $y,z$ plane, between the light propagation direction along $\hat{z}$ and the magnetic quantization axis. The top row $\theta=0$, is identical to the middle row ($\Delta m=0$) of Fig. \ref{fig:2}. Different patterns arise from different combinations of field gradient terms. For particular orientations, the excitations profiles for radially and azimuthally vortex light resembles the ones of HG beams. In particular, azimuthally polarized light leads to perfectly vanishing excitation strength for $\theta=0$, while the on-axis excitation is maximized for radially polarized light.}.
\label{fig:3}
\end{figure}

To exemplify the role of different light-matter interaction geometries, we consider the specific sub-transition $\Delta m=0$, for which the relative transition strength for different incidence angles are shown in Fig.~\ref{fig:3}. The electric field gradients appearing in equation~(\ref{eq:quadrupole-Hamiltonian}), mix under rotations and interfere, giving rise to complex spatial patterns of the transitions strengths. Interestingly, radially and azimuthally polarized LG beams, when driving quadrupole transition at certain angles with the magnetic field, affect the atom in a similar way that is produced when HG beams propagating along the quantization axis are impinging it. This can be understood by considering that in general LG beams can be written as superposition of HG beams. The different components in such superposition can be then modulated by the light-matter interaction geometry. Other useful features can be highlighted by quantifying the role of the incidence angle $\theta$ on the transitions strengths. For example, one can characterize the sensitivity on fluctuations on the direction of the magnetic field over a set of different $\Delta m$ quadrupole transitions. The effect of unavoidable magnetic field fluctuations can be then reduced by choosing the optimal configuration. More generally, the framework presented allows for exploring different geometries to tailor light-matter interaction for specific purposes. This could be particularly useful in the context of motional sidebands.

We emphasize that our formalism allows for the treatment of arbitrary structured light fields, beyond the HG and LG vector beams analysed in this work. We expect it to be of particular use when dealing with more complex electric field structures or topologies~\cite{sugic2021particle} and extremely focused beams, where no analytical description, but only numerical evaluation is available. In these cases, the framework will work with discretized electric field components and discretized gradients. For instance, numerical evaluation of focal fields \cite{novotny2012principles} produced via different real world optical systems can be used to carry out even more detailed predictions.

\section*{Coherent coupling of structured light fields to the motion of harmonically trapped atoms}

In this section, we extend the framework developed above to describe coherent coupling to the motion of single harmonically trapped ions or neutral atoms. We again focus on electric quadrupole transitions, which exhibit narrow linewidths, much smaller that the harmonic confinement secular frequencies. This admits optical driving within the \textit{resolved sideband regime} and therefore coherent driving of transitions between different oscillator eigenstates as achieved by weakly detuning the laser frequency. Note that some of the following predictions have been already experimentally tested~\cite{stopp2022coherent}. 

A Coulomb crystal of $N$ ions confined in a harmonic potential is described by a set of $3N$ collective modes of oscillations (normal modes). For strong confinement (resolved sideband regime) and low temperatures (Lamb-Dicke regime), quantum mechanical behaviour of the normal modes can be observed. The normal modes are independently quantized, using the annihilation and creation operators $\hat{a}_{i}$ and $\hat{a}^{\dagger}_{i}$ ($i=\{1,\cdots,3N\}$), which create or destroy motional quanta pertaining to mode $i$, respectively. In the following, we restrict the treatment to the case of a single ion, which has three secular modes. These describe oscillations at secular frequencies $\omega_q$ and along directions $q=\{X,Y,Z\}$, which are the mutually orthogonal main axes of the confining trap potential.
The oscillator eigenstates $\hat{a}^{\dagger}_q\hat{a}_q \ket{n_q}= n_q \ket{n_q}$ are characterized by motional quantum numbers $n_q$. In the resolved sideband regime, transitions between different eigenstates can be coherently driven. In order to describe the relative coupling strengths, we extend the framework introduced above by replacing the atomic center-of-mass coordinate $\vec{R}$ by $\vec{R}_0+\hat{\vec{R}}$, where $\vec{R}_0$ is the trap center and $\hat{\vec{R}}$ is the atomic position operators describing the displacement from the equilibrium position. The relative transition strength equation~(\ref{eq:relTransitionStrength}) then also becomes an operator $\hat{\mu}\left(\vec{R}_0+\hat{\vec{R}}\right)$. We compute the transition matrix elements between oscillator eigenstates with respect to the motional quantum numbers $\{n_q\}$:
\begin{equation} \label{eq:matrix-element-motional-excitations}
\mu(\vec{R}_0+\hat{\vec{R}},\{n_q'\},\{n_q\})=\braket{\{n_{q}'\}|\hat{\mu}\left(\vec{R}_0+\hat{\vec{R}}\right)|\{n_{q}\}}.
\end{equation}
The common treatment of motional transitions in the resolved sideband regime consists of assuming a plane-wave driving field, for which the electronic matrix elements do not depend on the atom position. The relative transition strength factorizes into the electronic matrix element and a motional matrix element, which is obtained by expressing the plane-wave field in terms of the atomic position operator $\hat{\vec{R}}$ and taking its matrix elements with respect to oscillator eigenstates \cite{Wineland1998}. The validity of this approach clearly breaks down for tightly focused drive fields.\\
First-order sideband transitions on mode $p$ are driven by detuning the drive field by $\pm \omega_p$ from the electronic (carrier) resonance. Upon transition from the initial to the final electronic state, the motional quantum number of the mode is changed by $\Delta n_p=\pm 1$. In the following, we describe how the relative transition amplitudes for first-order sideband transitions can be computed for arbitrary vector fields. To that end, we expand the bare (electronic) relative transition strength equation~(\ref{eq:relTransitionStrength}) around the trap center $\vec{R}_0$ up to first order in the atomic coordinates:
\begin{equation} \label{eq:Taylor-expansion}
\mu(\vec{R}_0+\vec{R}) = \mu(\vec{R}_0) \braket{\{n_q'\}|\{n_q\}}+ \sum_{q=X,Y,Z} \left. \frac{\partial \mu}{\partial q} \right|_{\vec{R}_0}  \braket{n_q'|\hat q|n_q} +\mathcal{O}(q^2).
\end{equation}
Restricting to a first-order sideband of mode $p$, i.e. $n_p'=n_p \pm 1$ and $n_q'=n_q$ for $q\neq p$, the zeroth order term vanishes and only one term from the first expansion order remains. Using $\hat{q}=\sqrt{\frac{\hbar}{2m\omega_q}}(\hat{a}_q+\hat{a}^{\dagger}_q)$, we derive the relative transition strengths for the blue sideband (bsb) $n_p'=n_p+1$ and red sideband (rsb) $n_p'=n_p-1$:
\begin{eqnarray} \label{eq:first-sidebands}
\mu_{bsb}(\vec{R}_0)&=&\left.\frac{\partial \mu}{\partial p}\right|_{\vec{R}_0}\sqrt{\frac{\hbar}{2m\omega_p}}\sqrt{n_p+1},\\\label{eq:second-sidebands}
\mu_{rsb}(\vec{R}_0)&=&\left.\frac{\partial \mu}{\partial p}\right|_{\vec{R}_0}\sqrt{\frac{\hbar}{2m\omega_p}}\sqrt{n_p}.
\end{eqnarray}
Considering again an electric quadrupole transition and assuming the special - but experimentally relevant - case that the beam axis system coincides with the trap axis system, we invoke equation~(\ref{eq:relTransitionStrength}) to calculate the spatial derivative of the bare transition strength:
\begin{equation}
    \partial_p\mu(\vec{R};J_2,J_1,m_2,m_2)\vert_{\vec{R}_0} = \sum_{i,j=X,Y,Z} \sum_{\Delta m=-\Delta J}^{+\Delta J} \braket{J_1 m_1 \Delta J \Delta m|J_2 m_2} c_{ij}^{(\Delta m)} \partial_p\partial_i E_j\vert_{\vec{R}_0}
    \label{eq:muderivative}
\end{equation}
For the case that the axes systems do not coincide, we have to employ the derivative direct along $\vec{e}_p$:
\begin{equation}
\partial_p= \sum_{q={X,Y,Z}} \left(\vec{e}_p\cdot\vec{e}_q\right) \partial_q
\end{equation}
Comparing equation~(\ref{eq:relTransitionStrength}) against equations~(\ref{eq:first-sidebands})~and~(\ref{eq:muderivative}), we see that for a quadrupole transition, the field gradients govern the bare electronic transition strengths, while second spatial derivatives of the field govern the coupling to the atomic motion. 

\begin{figure}[ht!]
\centering
\includegraphics[scale = 0.22]{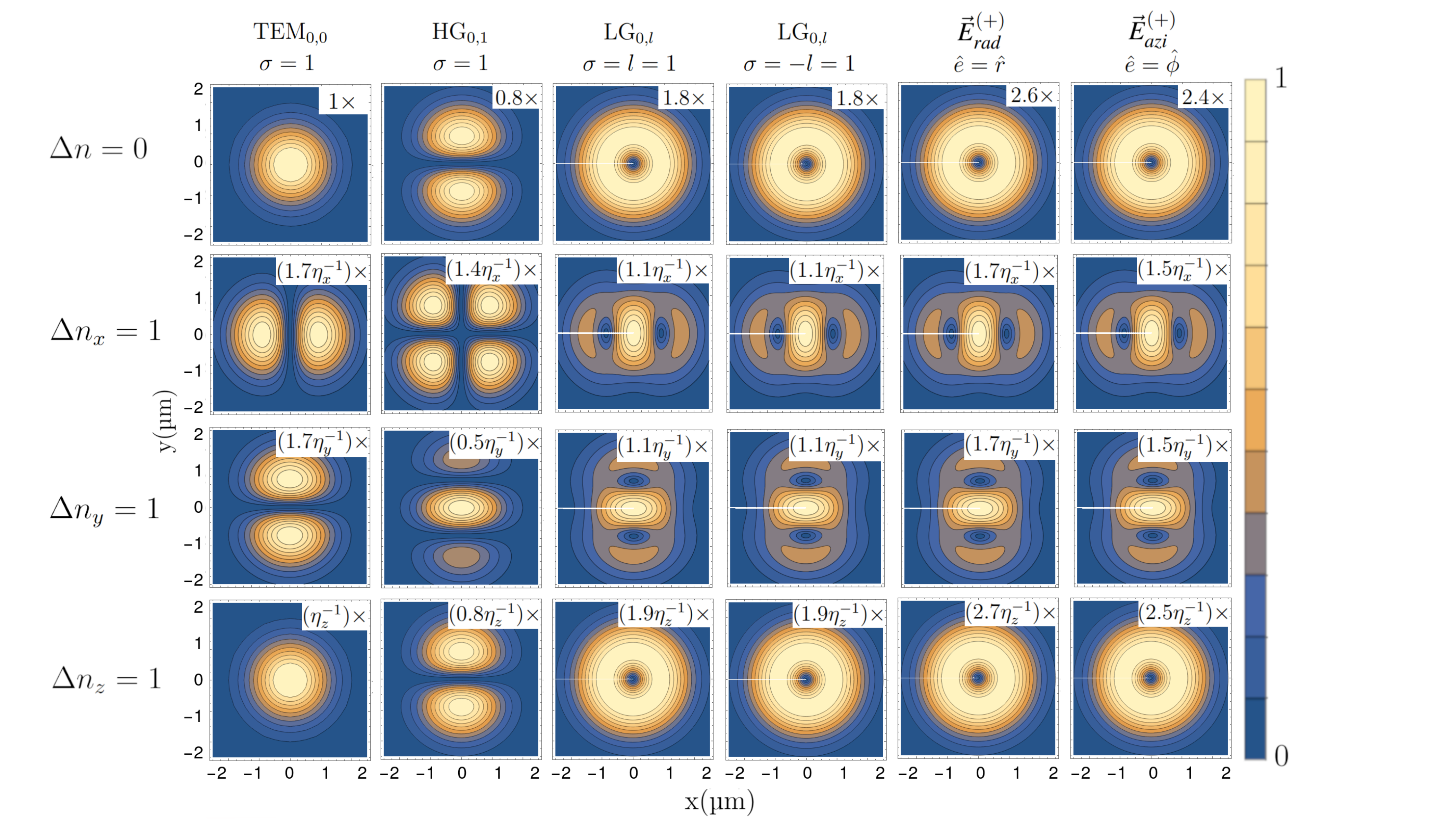}
\caption{Relative transition strengths for carrier $\Delta n=0$, see equation~(\ref{eq:relTransitionStrength}) and blue sideband $\Delta n_j=1$ transitions, see equation~(\ref{eq:first-sidebands}), are calculated with our \href{https://community.wolfram.com/groups/-/m/t/3099859}{Mathematica Notebook} for the $m=+\tfrac{1}{2} \leftrightarrow m=+\tfrac{3}{2}$ sub-transition of a $\Delta J=2$ quadrupole transition. Blue sideband strenghts are calculated under the assumption that the atom is in the motional ground state. The drive field propagates along the quantization axis $z$ with circular polarization $\sigma=+1$ (except for radially and azimuthally polarized beams), the trap axes are aligned with the laser axis system, and the beam types and parameters are the same as above. $\Delta n_j=1$ refers to the addition of a single motional quanta along the $j$-direction. The transition strengths for the longitudinal ($z$) and transversal ($x,y$) sideband, are rescaled by  the Lamb-Dicke parameters $\eta_z=k\sqrt{\frac{\hbar}{2m\omega_z}}$ and $\eta_{x,y}=\frac{\sqrt{2}}{w0}\sqrt{\frac{\hbar}{2m\omega_{x,y}}}$ introduced in Ref.~\cite{stopp2022coherent}.}
\label{fig:4}
\end{figure}

We apply this framework for a $\Delta J=2$ quadrupole transition to compute relative transition strengths for $\Delta n=0$ (carrier) and $\Delta n_q=1$ (blue sideband) transitions for all modes $q=X,Y,Z$, by assuming the atom being in its motional ground state and lying in the focal plane of the structured beams covered above. The extension to arbitrary sidebands with different intial motional state is straightforward by means of equations~(\ref{eq:first-sidebands}) and (\ref{eq:second-sidebands}). The results for magnetic sub-transitions $\Delta m=+ 1$ are shown in Fig.~\ref{fig:4}.  The trap axes coincide with the axis system defined by the laser beam,  with $z$ being the magnetic quantization axis as above. The laser beam also propagates along direction $z$.

As expected, the carrier excitation profiles ($\Delta n=0$), reflect the intensity profile of the $\sigma=\Delta m=+1$ polarization component, compare to first row of Fig.~\ref{fig:1}. Here, the  transition strength is governed by the optical phase gradient along the propagation direction, which is determined by the traveling wave term $e^{ikz}$. 
For the sidebands pertaining to modes along the transverse directions $x$ and $y$, we see that the transitions strengths are proportional to the gradients of the carrier transition strengths along the respective direction. For HG and LG beams, the transition strengths for these sidebands peak at the center of the beam. This is particularly interesting as at these points, the carrier excitation strength is vanishing, as has been experimentally shown in Ref. ~\cite{stopp2022coherent}. The bottom row of Fig.~\ref{fig:4} shows the sideband transitions strengths for the mode longitudinal to the beam propagation direction ($z$). As for the carrier transition, these are proportional to the beam intensity, as they are mainly generated by the traveling wave component $e^{ikz}$. Conversely (not shown), when $l\neq \sigma$ a distinct longitudinal field~\cite{Quinteiro2017} concentrates at the center of the focused beam, which changes the excitation profiles for the $\Delta m=0$ transition, see Fig.~\ref{fig:2}.

Note that the framework introduced above can be used to predict the transversal motional excitations of a single trapped ion when is placed in the phase singularity of a vortex beam or within a slightly misaligned Gaussian beam, as respectively reported in \cite{stopp2022coherent} and \cite{West_2021}. In both works, the electric field gradients perpendicular to the light propagation direction lead to transversal motion excitations,  where the transition strengths are described by an effective transverse Lamb-Dicke parameter $\eta_{\perp}$ and modulated by the beam mode profiles and polarization structures.

\section*{Discussion and outlook}
This work provides a unified and generalized treatment of light-matter interaction for strongly localized atoms interacting with structured light fields, beyond the paraxial approximation. We explicitly derived equations describing atomic dipole and quadrupole transitions, for arbitrary interaction geometry as defined by the light propagation direction and the quantizing magnetic field. The treatment was extended to describe the coupling to the atomic motion for the case of harmonically trapped atoms, with a specific focus on trapped-ion experiments. In this context, we highlighted how the introduced formalism precisely describes the surprising predictions, reported in literature, when trapped ions interact with vortex light fields driving electronic quadrupole transitions  and their longitudinal and transversal motional sidebands. \\
Our generalized framework allows for the quantitative description of light-matter interaction using more general vector fields with arbitrary mode functions and spatially modulated polarization. In conjunction with the provided code, it allows for exploring a wealth of phenomena occurring when atomic systems are coherently manipulated using structured light fields. In the following, we discuss a number of possible application scenarios.\\
As noted initially, technological applications for structured light fields currently emerge, including e.g. photolithography \cite{li2018}. In the context of such applications, trapped atoms may serve as precise probes with sub-wavelength resolution for mapping out the properties of the light fields, in order to characterize and improve optical instrumentation.\\
Structured light fields can be a valuable asset for many emerging quantum technologies. For example, optical frequency standards~\cite{PhysRevLett.129.253901} can be improved using vortex beams, allowing for ac Stark-shift-free excitation of clock transitions. In the context of trapped-ion quantum computing, vortex beams can be applied to realize optimized quantum gates with minimized parasitic Stark shifts and off-resonant scattering. In particular, vortex beams can efficiently generate state-dependent optical forces transverse to the beam propagation direction \cite{mazzanti2023trapped}, which can allow for increased versatility of optical addressing schemes operating on linear Coulomb crystals. For many trapped-ion qubit realizations, single- and multi-qubit gates are realized via two-photon Raman transitions mediated by off-resonant E1 electric dipole transitions. For example, for qubits based on the spin of the valence electron of $^{40}$Ca$^+$ ions~\cite{Poschinger_2009,kaushal2020shuttling,hilder2022fault}, stimulated Raman transitions mediated by the $S_{1/2} \leftrightarrow P_{1/2}$ transition coherently drive transitions between the Zeeman sublevels $m_J=\pm 1/2$ of the electronic ground state and the respective motional sidebands. Using such stimulated Raman transitions in conjunction with structured light fields may lead to interesting applications, for example to drive motional sideband without parasitic carrier excitation. Moreover,  spin-dependent optical forces generated by vortex beams may be used to implement coherent spin-dependent ion SWAP operations~\cite{kaufmann2017fast,urban2019coherent}. Finally, highly excited Rydberg atoms or ions~\cite{rodrigues2016excitation,Niederlaender2023} can provide the possibility to study light-matter interaction on the single atom level in a new regime, where the spatial features of the driving fields are smaller or of the same order than the spatial extent of the electronic wavefunctions.

\bibliography{main}

\begin{thebibliography}{72}
\providecommand{\natexlab}[1]{#1}
\providecommand{\url}[1]{\texttt{#1}}
\expandafter\ifx\csname urlstyle\endcsname\relax
  \providecommand{\doi}[1]{doi: #1}\else
  \providecommand{\doi}{doi: \begingroup \urlstyle{rm}\Url}\fi

\bibitem[Kuga et~al.(1997)Kuga, Torii, Shiokawa, Hirano, Shimizu, and Sasada]{kuga1997novel}
Takahiro Kuga, Yoshio Torii, Noritsugu Shiokawa, Takuya Hirano, Yukiko Shimizu, and Hiroyuki Sasada.
\newblock Novel optical trap of atoms with a doughnut beam.
\newblock \emph{Phys. Rev. Lett.}, 78\penalty0 (25):\penalty0 4713, 1997.

\bibitem[He et~al.(1995)He, Friese, Heckenberg, and Rubinsztein-Dunlop]{he1995direct}
H~He, MEJ Friese, NR~Heckenberg, and H~Rubinsztein-Dunlop.
\newblock Direct observation of transfer of angular momentum to absorptive particles from a laser beam with a phase singularity.
\newblock \emph{Phys. Rev. Lett.}, 75\penalty0 (5):\penalty0 826, 1995.

\bibitem[Padgett and Bowman(2011)]{padgett2011tweezers}
Miles Padgett and Richard Bowman.
\newblock Tweezers with a twist.
\newblock \emph{Nat. Photon.}, 5\penalty0 (6):\penalty0 343, 2011.

\bibitem[Stilgoe et~al.(2022)Stilgoe, Nieminen, and Rubinsztein-Dunlop]{stilgoe2022controlled}
Alexander~B Stilgoe, Timo~A Nieminen, and Halina Rubinsztein-Dunlop.
\newblock Controlled transfer of transverse orbital angular momentum to optically trapped birefringent microparticles.
\newblock \emph{Nat. Photon.}, 16\penalty0 (5):\penalty0 346, 2022.

\bibitem[Hell and Wichmann(1994)]{hell1994breaking}
Stefan~W Hell and Jan Wichmann.
\newblock Breaking the diffraction resolution limit by stimulated emission: stimulated-emission-depletion fluorescence microscopy.
\newblock \emph{Opt. Lett.}, 19\penalty0 (11):\penalty0 780, 1994.

\bibitem[Swartzlander(2001)]{swartzlander2001peering}
Grover~A Swartzlander.
\newblock Peering into darkness with a vortex spatial filter.
\newblock \emph{Opt. Lett.}, 26\penalty0 (8):\penalty0 497, 2001.

\bibitem[F{\"u}rhapter et~al.(2005)F{\"u}rhapter, Jesacher, Bernet, and Ritsch-Marte]{furhapter2005spiral}
Severin F{\"u}rhapter, Alexander Jesacher, Stefan Bernet, and Monika Ritsch-Marte.
\newblock Spiral interferometry.
\newblock \emph{Opt. Lett.}, 30\penalty0 (15):\penalty0 1953, 2005.

\bibitem[Sanner et~al.(2019)Sanner, Huntemann, Lange, Tamm, Peik, Safronova, and Porsev]{sanner2019optical}
Christian Sanner, Nils Huntemann, Richard Lange, Christian Tamm, Ekkehard Peik, Marianna~S Safronova, and Sergey~G Porsev.
\newblock Optical clock comparison for lorentz symmetry testing.
\newblock \emph{Nature}, 567\penalty0 (7747):\penalty0 204, 2019.

\bibitem[Lange et~al.(2021)Lange, Huntemann, Rahm, Sanner, Shao, Lipphardt, Tamm, Weyers, and Peik]{lange2021improved}
Richard Lange, Nils Huntemann, JM~Rahm, Christian Sanner, Hu~Shao, Burghard Lipphardt, Chr Tamm, Stefan Weyers, and Ekkehard Peik.
\newblock Improved limits for violations of local position invariance from atomic clock comparisons.
\newblock \emph{Phys. Rev. Lett.}, 126\penalty0 (1):\penalty0 011102, 2021.

\bibitem[H{\"a}ffner et~al.(2008)H{\"a}ffner, Roos, and Blatt]{haffner2008quantum}
Hartmut H{\"a}ffner, Christian~F Roos, and Rainer Blatt.
\newblock Quantum computing with trapped ions.
\newblock \emph{Phys. Rep.}, 469\penalty0 (4):\penalty0 155, 2008.

\bibitem[Henriet et~al.(2020)Henriet, Beguin, Signoles, Lahaye, Browaeys, Reymond, and Jurczak]{henriet2020quantum}
Lo{\"\i}c Henriet, Lucas Beguin, Adrien Signoles, Thierry Lahaye, Antoine Browaeys, Georges-Olivier Reymond, and Christophe Jurczak.
\newblock Quantum computing with neutral atoms.
\newblock \emph{Quantum}, 4:\penalty0 327, 2020.

\bibitem[Schmiegelow and Schmidt-Kaler(2012)]{schmiegelow2012light}
Christian~Tom{\'a}s Schmiegelow and Ferdinand Schmidt-Kaler.
\newblock Light with orbital angular momentum interacting with trapped ions.
\newblock \emph{Eur. Phys. J. D}, 66\penalty0 (6):\penalty0 1, 2012.

\bibitem[Solyanik-Gorgone et~al.(2019)Solyanik-Gorgone, Afanasev, Carlson, Schmiegelow, and Schmidt-Kaler]{Solyanik2019}
Maria Solyanik-Gorgone, Andrei Afanasev, Carl~E. Carlson, Christian~T. Schmiegelow, and Ferdinand Schmidt-Kaler.
\newblock Excitation of ${E}$1-forbidden atomic transitions with electric, magnetic, or mixed multipolarity in light fields carrying orbital and spin angular momentum.
\newblock \emph{J. Opt. Soc. Am. B}, 36\penalty0 (3):\penalty0 565, Mar 2019.

\bibitem[Peshkov et~al.(2023{\natexlab{a}})Peshkov, Bidasyuk, Lange, Huntemann, Peik, and Surzhykov]{PhysRevA.107.023106}
A.~A. Peshkov, Y.~M. Bidasyuk, R.~Lange, N.~Huntemann, E.~Peik, and A.~Surzhykov.
\newblock Interaction of twisted light with a trapped atom: Interplay between electronic and motional degrees of freedom.
\newblock \emph{Phys. Rev. A}, 107:\penalty0 023106, Feb 2023{\natexlab{a}}.

\bibitem[Padgett(2017)]{padgett2017orbital}
Miles~J Padgett.
\newblock Orbital angular momentum 25 years on.
\newblock \emph{Opt. Express}, 25\penalty0 (10):\penalty0 11265, 2017.

\bibitem[Quabis et~al.(2000)Quabis, Dorn, Eberler, Glöckl, and Leuchs]{QUABIS20001}
S~Quabis, R~Dorn, M~Eberler, O~Glöckl, and G~Leuchs.
\newblock Focusing light to a tighter spot.
\newblock \emph{Opt. Commun.}, 179\penalty0 (1):\penalty0 1--7, 2000.
\newblock ISSN 0030-4018.

\bibitem[Monteiro et~al.(2009{\natexlab{a}})Monteiro, Neto, and Nussenzveig]{PhysRevA.79.033830}
Paula~B. Monteiro, Paulo A.~Maia Neto, and H.~Moys\'es Nussenzveig.
\newblock Angular momentum of focused beams: Beyond the paraxial approximation.
\newblock \emph{Phys. Rev. A}, 79:\penalty0 033830, Mar 2009{\natexlab{a}}.

\bibitem[Alber et~al.(2017)Alber, Fischer, Bader, Mantel, Sondermann, and Leuchs]{alber2017focusing}
Lucas Alber, Martin Fischer, Marianne Bader, Klaus Mantel, Markus Sondermann, and Gerd Leuchs.
\newblock Focusing characteristics of a 4 $\pi$ parabolic mirror light-matter interface.
\newblock \emph{Journal of the European Optical Society-Rapid Publications}, 13:\penalty0 1, 2017.

\bibitem[Araneda et~al.(2020)Araneda, Cerchiari, Higginbottom, Holz, Lakhmanskiy, Ob{\v{s}}il, Colombe, and Blatt]{araneda2020panopticon}
Gabriel Araneda, Giovanni Cerchiari, Daniel~B Higginbottom, Philip~C Holz, Kirill Lakhmanskiy, Petr Ob{\v{s}}il, Yves Colombe, and Rainer Blatt.
\newblock The panopticon device: An integrated paul-trap--hemispherical mirror system for quantum optics.
\newblock \emph{Rev. Sci. Instrum.}, 91\penalty0 (11):\penalty0 113201, 2020.

\bibitem[Jefferts et~al.(1995)Jefferts, Monroe, Barton, and Wineland]{jefferts1995paul}
Steven~R Jefferts, C~Monroe, AS~Barton, and David~J Wineland.
\newblock Paul trap for optical frequency standards.
\newblock \emph{IEEE Trans. Instrum. Meas.}, 44\penalty0 (2):\penalty0 148, 1995.

\bibitem[Blaum et~al.(2020)Blaum, Eliseev, and Sturm]{blaum2020perspectives}
Klaus Blaum, Sergey Eliseev, and Sven Sturm.
\newblock Perspectives on testing fundamental physics with highly charged ions in penning traps.
\newblock \emph{Quantum Science and Technology}, 6\penalty0 (1):\penalty0 014002, 2020.

\bibitem[Kaufman et~al.(2012)Kaufman, Lester, and Regal]{kaufman2012cooling}
Adam~M Kaufman, Brian~J Lester, and Cindy~A Regal.
\newblock Cooling a single atom in an optical tweezer to its quantum ground state.
\newblock \emph{Physical Review X}, 2\penalty0 (4):\penalty0 041014, 2012.

\bibitem[Kolesov et~al.(2012)Kolesov, Xia, Reuter, St{\"o}hr, Zappe, Meijer, Hemmer, and Wrachtrup]{kolesov2012optical}
R~Kolesov, K~Xia, R~Reuter, R~St{\"o}hr, A~Zappe, J~Meijer, PR~Hemmer, and J~Wrachtrup.
\newblock Optical detection of a single rare-earth ion in a crystal.
\newblock \emph{Nat. Commun.}, 3\penalty0 (1):\penalty0 1029, 2012.

\bibitem[Bradac et~al.(2019)Bradac, Gao, Forneris, Trusheim, and Aharonovich]{bradac2019quantum}
Carlo Bradac, Weibo Gao, Jacopo Forneris, Matthew~E Trusheim, and Igor Aharonovich.
\newblock Quantum nanophotonics with group iv defects in diamond.
\newblock \emph{Nat. Commun.}, 10\penalty0 (1):\penalty0 5625, 2019.

\bibitem[Schmiegelow et~al.(2016)Schmiegelow, Schulz, Kaufmann, Ruster, Poschinger, and Schmidt-Kaler]{schmiegelow2016transfer}
Christian~T Schmiegelow, Jonas Schulz, Henning Kaufmann, Thomas Ruster, Ulrich~G Poschinger, and Ferdinand Schmidt-Kaler.
\newblock Transfer of optical orbital angular momentum to a bound electron.
\newblock \emph{Nat. Commun.}, 7\penalty0 (1):\penalty0 12998, 2016.

\bibitem[Drechsler et~al.(2021)Drechsler, Wolf, Schmiegelow, and Schmidt-Kaler]{drechsler2021optical}
Mart{\'\i}n Drechsler, Sebastian Wolf, Christian~T Schmiegelow, and Ferdinand Schmidt-Kaler.
\newblock Optical superresolution sensing of a trapped ion’s wave packet size.
\newblock \emph{Phys. Rev. Lett.}, 127\penalty0 (14):\penalty0 143602, 2021.

\bibitem[Stopp et~al.(2022)Stopp, Verde, Katz, Drechsler, Schmiegelow, and Schmidt-Kaler]{stopp2022coherent}
Felix Stopp, Maurizio Verde, Milton Katz, Martin Drechsler, Christian~T. Schmiegelow, and Ferdinand Schmidt-Kaler.
\newblock Coherent transfer of transverse optical momentum to the motion of a single trapped ion.
\newblock \emph{Phys. Rev. Lett.}, 129:\penalty0 263603, Dec 2022.

\bibitem[Quinteiro et~al.(2017{\natexlab{a}})Quinteiro, Schmidt-Kaler, and Schmiegelow]{Quinteiro2017}
G.~F. Quinteiro, Ferdinand Schmidt-Kaler, and Christian~T. Schmiegelow.
\newblock Twisted-light--ion interaction: The role of longitudinal fields.
\newblock \emph{Phys. Rev. Lett.}, 119:\penalty0 253203, Dec 2017{\natexlab{a}}.

\bibitem[Afanasev et~al.(2018)Afanasev, Carlson, Schmiegelow, Schulz, Schmidt-Kaler, and Solyanik]{Afanasev_2018}
Andrei Afanasev, Carl~E Carlson, Christian~T Schmiegelow, Jonas Schulz, Ferdinand Schmidt-Kaler, and Maria Solyanik.
\newblock Experimental verification of position-dependent angular-momentum selection rules for absorption of twisted light by a bound electron.
\newblock \emph{New Journal of Physics}, 20\penalty0 (2):\penalty0 023032, feb 2018.

\bibitem[Qian et~al.(2021)Qian, Cui, Luo, Zheng, Huang, Ai, He, Li, and Guo]{PhysRevLett.127.263603}
Zhong-Hua Qian, Jin-Ming Cui, Xi-Wang Luo, Yong-Xiang Zheng, Yun-Feng Huang, Ming-Zhong Ai, Ran He, Chuan-Feng Li, and Guang-Can Guo.
\newblock Super-resolved imaging of a single cold atom on a nanosecond timescale.
\newblock \emph{Phys. Rev. Lett.}, 127:\penalty0 263603, Dec 2021.

\bibitem[Lange et~al.(2022)Lange, Huntemann, Peshkov, Surzhykov, and Peik]{PhysRevLett.129.253901}
R.~Lange, N.~Huntemann, A.~A. Peshkov, A.~Surzhykov, and E.~Peik.
\newblock Excitation of an electric octupole transition by twisted light.
\newblock \emph{Phys. Rev. Lett.}, 129:\penalty0 253901, Dec 2022.

\bibitem[Lax et~al.(1975)Lax, Louisell, and McKnight]{lax1975maxwell}
Melvin Lax, William~H Louisell, and William~B McKnight.
\newblock From maxwell to paraxial wave optics.
\newblock \emph{Physical Review A}, 11\penalty0 (4):\penalty0 1365, 1975.

\bibitem[Siegman(1986)]{siegman1986lasers}
AE~Siegman.
\newblock Lasers (mill valley, ca: University science).
\newblock \emph{Chap}, 17:\penalty0 663, 1986.

\bibitem[Barnett(2001)]{barnett2001optical}
Stephen~M Barnett.
\newblock Optical angular-momentum flux.
\newblock \emph{J. Opt. B: Quantum Semiclass. Opt.}, 4\penalty0 (2):\penalty0 S7, 2001.

\bibitem[Van~Enk and Nienhuis(1994{\natexlab{a}})]{van1994commutation}
SJ~Van~Enk and G~Nienhuis.
\newblock Commutation rules and eigenvalues of spin and orbital angular momentum of radiation fields.
\newblock \emph{J. Mod. Opt.}, 41\penalty0 (5):\penalty0 963, 1994{\natexlab{a}}.

\bibitem[Van~Enk and Nienhuis(1994{\natexlab{b}})]{van1994spin}
SJ~Van~Enk and G~Nienhuis.
\newblock Spin and orbital angular momentum of photons.
\newblock \emph{Europhys. Lett.}, 25\penalty0 (7):\penalty0 497, 1994{\natexlab{b}}.

\bibitem[Bliokh et~al.(2015)Bliokh, Rodr{\'\i}guez-Fortu{\~n}o, Nori, and Zayats]{bliokh2015spin}
Konstantin~Yu Bliokh, Francisco~J Rodr{\'\i}guez-Fortu{\~n}o, Franco Nori, and Anatoly~V Zayats.
\newblock Spin--orbit interactions of light.
\newblock \emph{Nat. Photon.}, 9\penalty0 (12):\penalty0 796, 2015.

\bibitem[Barnett and Allen(1994)]{barnett1994orbital}
Stephen~M Barnett and L~Allen.
\newblock Orbital angular momentum and nonparaxial light beams.
\newblock \emph{Opt. Commun.}, 110\penalty0 (5-6):\penalty0 670, 1994.

\bibitem[Takenaka et~al.(1985)Takenaka, Yokota, and Fukumitsu]{takenaka1985propagation}
Takashi Takenaka, Mitsuhiro Yokota, and Otozo Fukumitsu.
\newblock Propagation of light beams beyond the paraxial approximation.
\newblock \emph{J. Opt. Soc. Am. A}, 2\penalty0 (6):\penalty0 826, 1985.

\bibitem[Novotny and Hecht(2012)]{novotny2012principles}
Lukas Novotny and Bert Hecht.
\newblock \emph{Principles of nano-optics}.
\newblock Cambridge university press, 2012.

\bibitem[Dorn et~al.(2003)Dorn, Quabis, and Leuchs]{dorn2003focus}
Ralf Dorn, Susanne Quabis, and Gerd Leuchs.
\newblock The focus of light—linear polarization breaks the rotational symmetry of the focal spot.
\newblock \emph{J. Mod. Opt.}, 50\penalty0 (12):\penalty0 1917, 2003.

\bibitem[Zhan(2013)]{zhan2013vectorial}
Qiwen Zhan.
\newblock \emph{Vectorial optical fields: Fundamentals and applications}.
\newblock World scientific, 2013.

\bibitem[Quinteiro et~al.(2015)Quinteiro, Reiter, and Kuhn]{quinteiro2015formulation}
Guillermo~Federico Quinteiro, DE~Reiter, and T~Kuhn.
\newblock Formulation of the twisted-light--matter interaction at the phase singularity: The twisted-light gauge.
\newblock \emph{Physical Review A}, 91\penalty0 (3):\penalty0 033808, 2015.

\bibitem[Quinteiro et~al.(2017{\natexlab{b}})Quinteiro, Reiter, and Kuhn]{quinteiro2017formulation}
Guillermo~Federico Quinteiro, DE~Reiter, and T~Kuhn.
\newblock Formulation of the twisted-light--matter interaction at the phase singularity: beams with strong magnetic fields.
\newblock \emph{Physical Review A}, 95\penalty0 (1):\penalty0 012106, 2017{\natexlab{b}}.

\bibitem[Richards and Wolf(1959)]{richards1959electromagnetic}
Bernard Richards and Emil Wolf.
\newblock Electromagnetic diffraction in optical systems, ii. structure of the image field in an aplanatic system.
\newblock \emph{Proc. R. Soc. A: Math. Phys. Sci.}, 253\penalty0 (1274):\penalty0 358--379, 1959.

\bibitem[Chen et~al.(2002)Chen, Konkola, Ferrera, Heilmann, and Schattenburg]{chen2002analyses}
Carl~G Chen, Paul~T Konkola, Juan Ferrera, Ralf~K Heilmann, and Mark~L Schattenburg.
\newblock Analyses of vector gaussian beam propagation and the validity of paraxial and spherical approximations.
\newblock \emph{J. Opt. Soc. Am. A}, 19\penalty0 (2):\penalty0 404, 2002.

\bibitem[Wang et~al.(2020)Wang, Castellucci, and Franke-Arnold]{wang2020vectorial}
Jinwen Wang, Francesco Castellucci, and Sonja Franke-Arnold.
\newblock Vectorial light--matter interaction: Exploring spatially structured complex light fields.
\newblock \emph{AVS Quantum Science}, 2\penalty0 (3):\penalty0 031702, 2020.

\bibitem[Huang et~al.(2011)Huang, Shi, Cao, Li, Zhang, and Li]{huang2011vector}
Kun Huang, Peng Shi, GW~Cao, Ke~Li, XB~Zhang, and YP~Li.
\newblock Vector-vortex bessel--gauss beams and their tightly focusing properties.
\newblock \emph{Opt. Lett.}, 36\penalty0 (6):\penalty0 888, 2011.

\bibitem[Forbes et~al.(2021)Forbes, Green, and Jones]{forbes2021relevance}
Kayn~A Forbes, Dale Green, and Garth~A Jones.
\newblock Relevance of longitudinal fields of paraxial optical vortices.
\newblock \emph{Journal of Optics}, 23\penalty0 (7):\penalty0 075401, 2021.

\bibitem[Monteiro et~al.(2009{\natexlab{b}})Monteiro, Neto, and Nussenzveig]{monteiro2009angular}
Paula~B Monteiro, Paulo A~Maia Neto, and H~Moys{\'e}s Nussenzveig.
\newblock Angular momentum of focused beams: Beyond the paraxial approximation.
\newblock \emph{Physical Review A}, 79\penalty0 (3):\penalty0 033830, 2009{\natexlab{b}}.

\bibitem[Davis(1979)]{davis1979theory}
LW~Davis.
\newblock Theory of electromagnetic beams.
\newblock \emph{Phys. Rev. A}, 19\penalty0 (3):\penalty0 1177, 1979.

\bibitem[Maurer et~al.(2007)Maurer, Jesacher, F{\"u}rhapter, Bernet, and Ritsch-Marte]{maurer2007tailoring}
Christian Maurer, Alexander Jesacher, Severin F{\"u}rhapter, Stefan Bernet, and Monika Ritsch-Marte.
\newblock Tailoring of arbitrary optical vector beams.
\newblock \emph{New Journal of Physics}, 9\penalty0 (3):\penalty0 78, 2007.

\bibitem[Van~Enk(1994)]{van1994selection}
SJ~Van~Enk.
\newblock Selection rules and centre-of-mass motion of ultracold atoms.
\newblock \emph{Quantum Optics: Journal of the European Optical Society Part B}, 6\penalty0 (5):\penalty0 445, 1994.

\bibitem[Sonnleitner and Barnett(2017)]{sonnleitner2017rontgen}
Matthias Sonnleitner and Stephen~M Barnett.
\newblock The r{\"o}ntgen interaction and forces on dipoles in time-modulated optical fields.
\newblock \emph{Eur. Phys. J. D}, 71:\penalty0 1, 2017.

\bibitem[Peshkov et~al.(2023{\natexlab{b}})Peshkov, Jordan, Kromrey, Mehta, Mehlst{\"a}ubler, and Surzhykov]{peshkov2023excitation}
Anton~A Peshkov, Elena Jordan, Markus Kromrey, Karan~K Mehta, Tanja~E Mehlst{\"a}ubler, and Andrey Surzhykov.
\newblock Excitation of forbidden electronic transitions in atoms by hermite-gaussian modes.
\newblock \emph{Preprint at https://arxiv.org/pdf/2305.04523.pdf}, 2023{\natexlab{b}}.

\bibitem[Sakurai(1994)]{sakurai1994modern}
JJ~Sakurai.
\newblock Modern quantum mechanics (revised edition) addison-wesley.
\newblock \emph{Reading}, 41:\penalty0 221--223, 1994.

\bibitem[Afanasev et~al.(2013)Afanasev, Carlson, and Mukherjee]{afanasev2013off}
Andrei Afanasev, Carl~E Carlson, and Asmita Mukherjee.
\newblock Off-axis excitation of hydrogenlike atoms by twisted photons.
\newblock \emph{Physical Review A}, 88\penalty0 (3):\penalty0 033841, 2013.

\bibitem[Afanasev et~al.(2016)Afanasev, Carlson, and Mukherjee]{afanasev2016high}
Andrei Afanasev, Carl~E Carlson, and Asmita Mukherjee.
\newblock High-multipole excitations of hydrogen-like atoms by twisted photons near a phase singularity.
\newblock \emph{Journal of Optics}, 18\penalty0 (7):\penalty0 074013, 2016.

\bibitem[Afanasev et~al.(2020)Afanasev, Carlson, and Wang]{afanasev2020polarization}
Andrei Afanasev, Carl~E Carlson, and Hao Wang.
\newblock Polarization transfer from the twisted light to an atom.
\newblock \emph{Journal of Optics}, 22\penalty0 (5):\penalty0 054001, 2020.

\bibitem[Al-Khateeb et~al.(2022)Al-Khateeb, Lyras, Lembessis, and Aldossary]{al2022two}
Ayman Al-Khateeb, Andreas Lyras, VE~Lembessis, and Omar~M Aldossary.
\newblock Two-photon bound--bound atomic transitions induced by lg beams.
\newblock \emph{Results Phys.}, 43:\penalty0 106107, 2022.

\bibitem[Sugic et~al.(2021)Sugic, Droop, Otte, Ehrmanntraut, Nori, Ruostekoski, Denz, and Dennis]{sugic2021particle}
Danica Sugic, Ramon Droop, Eileen Otte, Daniel Ehrmanntraut, Franco Nori, Janne Ruostekoski, Cornelia Denz, and Mark~R Dennis.
\newblock Particle-like topologies in light.
\newblock \emph{Nat. Commun.}, 12\penalty0 (1):\penalty0 6785, 2021.

\bibitem[Wineland et~al.(1998)Wineland, Monroe, Itano, Leibfried, King, and Meekhof]{Wineland1998}
D.~J. Wineland, C.~Monroe, W.~M. Itano, D.~Leibfried, B.~E. King, and D.~M. Meekhof.
\newblock Experimental issues in coherent quantum-state manipulation of trapped atomic ions.
\newblock \emph{J. Res. Natl. Inst. Stand. Technol.}, 103:\penalty0 259, 1998.
\newblock URL \url{https://tf.nist.gov/general/pdf/1275.pdf}.

\bibitem[West et~al.(2021)West, Putnam, Campbell, and Hamilton]{West_2021}
Adam~D West, Randall Putnam, Wesley~C Campbell, and Paul Hamilton.
\newblock Tunable transverse spin–motion coupling for quantum information processing.
\newblock \emph{Quantum Science and Technology}, 6\penalty0 (2):\penalty0 024003, jan 2021.

\bibitem[Li et~al.(2018)Li, Liang, Zhan, Xu, Bai, and Wang]{li2018}
Xiongfeng Li, Yiyong Liang, Shichao Zhan, Jianfeng Xu, Jian Bai, and Kaiwei Wang.
\newblock Optical vortex beam direct-writing photolithography.
\newblock \emph{Applied Physics Express}, 11\penalty0 (3):\penalty0 036503, feb 2018.

\bibitem[Mazzanti et~al.(2023)Mazzanti, Gerritsma, Spreeuw, and Safavi-Naini]{mazzanti2023trapped}
M~Mazzanti, R~Gerritsma, RJC Spreeuw, and A~Safavi-Naini.
\newblock Trapped ion quantum computing using optical tweezers and the magnus effect.
\newblock \emph{arXiv preprint arXiv:2301.04668}, 2023.

\bibitem[Poschinger et~al.(2009)Poschinger, Huber, Ziesel, Deiß, Hettrich, Schulz, Singer, Poulsen, Drewsen, Hendricks, and Schmidt-Kaler]{Poschinger_2009}
U~G Poschinger, G~Huber, F~Ziesel, M~Deiß, M~Hettrich, S~A Schulz, K~Singer, G~Poulsen, M~Drewsen, R~J Hendricks, and F~Schmidt-Kaler.
\newblock Coherent manipulation of a 40ca+ spin qubit in a micro ion trap.
\newblock \emph{Journal of Physics B: Atomic, Molecular and Optical Physics}, 42\penalty0 (15):\penalty0 154013, jul 2009.

\bibitem[Kaushal et~al.(2020)Kaushal, Lekitsch, Stahl, Hilder, Pijn, Schmiegelow, Bermudez, M{\"u}ller, Schmidt-Kaler, and Poschinger]{kaushal2020shuttling}
V~Kaushal, Bjoern Lekitsch, A~Stahl, J~Hilder, Daniel Pijn, C~Schmiegelow, Alejandro Bermudez, M~M{\"u}ller, Ferdinand Schmidt-Kaler, and U~Poschinger.
\newblock Shuttling-based trapped-ion quantum information processing.
\newblock \emph{AVS Quantum Science}, 2\penalty0 (1):\penalty0 014101, 2020.

\bibitem[Hilder et~al.(2022)Hilder, Pijn, Onishchenko, Stahl, Orth, Lekitsch, Rodriguez-Blanco, M{\"u}ller, Schmidt-Kaler, and Poschinger]{hilder2022fault}
Janine Hilder, Daniel Pijn, Oleksiy Onishchenko, Alexander Stahl, Maximilian Orth, Bj{\"o}rn Lekitsch, Andrea Rodriguez-Blanco, Markus M{\"u}ller, Ferdinand Schmidt-Kaler, and UG~Poschinger.
\newblock Fault-tolerant parity readout on a shuttling-based trapped-ion quantum computer.
\newblock \emph{Physical Review X}, 12\penalty0 (1):\penalty0 011032, 2022.

\bibitem[Kaufmann et~al.(2017)Kaufmann, Ruster, Schmiegelow, Luda, Kaushal, Schulz, von Lindenfels, Schmidt-Kaler, and Poschinger]{kaufmann2017fast}
Henning Kaufmann, Thomas Ruster, Christian~T Schmiegelow, Marcelo~A Luda, Vidyut Kaushal, Jonas Schulz, David von Lindenfels, Ferdinand Schmidt-Kaler, and Ulrich~G Poschinger.
\newblock Fast ion swapping for quantum-information processing.
\newblock \emph{Physical Review A}, 95\penalty0 (5):\penalty0 052319, 2017.

\bibitem[Urban et~al.(2019)Urban, Glikin, Mouradian, Krimmel, Hemmerling, and Haeffner]{urban2019coherent}
Erik Urban, Neil Glikin, Sara Mouradian, Kai Krimmel, Boerge Hemmerling, and Hartmut Haeffner.
\newblock Coherent control of the rotational degree of freedom of a two-ion coulomb crystal.
\newblock \emph{Phys. Rev. Lett.}, 123\penalty0 (13):\penalty0 133202, 2019.

\bibitem[Rodrigues et~al.(2016)Rodrigues, Marcassa, and Mendon{\c{c}}a]{rodrigues2016excitation}
JD~Rodrigues, Luis~Gustavo Marcassa, and JT~Mendon{\c{c}}a.
\newblock Excitation of high orbital angular momentum rydberg states with laguerre--gauss beams.
\newblock \emph{J. Phys. B: At. Mol. Opt. Phys.}, 49\penalty0 (7):\penalty0 074007, 2016.

\bibitem[Niederländer et~al.(2023)Niederländer, Vogel, Schulze-Makuch, Gély, Mokhberi, and Schmidt-Kaler]{Niederlaender2023}
Marie Niederländer, Jonas Vogel, Alexander Schulze-Makuch, Bastien Gély, Arezoo Mokhberi, and Ferdinand Schmidt-Kaler.
\newblock Rydberg ions in coherent motional states: a new method for determining the polarizability of rydberg ions.
\newblock \emph{New Journal of Physics}, 25\penalty0 (3):\penalty0 033020, mar 2023.

\end{thebibliography}
\end{document}